\documentclass[preprintnumbers, prd, onecolumn, showpacs,floatfix,preprintnumbers,
superscriptaddress, nofootinbib]{revtex4}
\usepackage{graphicx}
\usepackage{epsfig}
\usepackage{bm}
\usepackage{amssymb}
\usepackage{float}
\usepackage{amsmath}
\usepackage{subfigure}
\usepackage{dcolumn}
\usepackage{cancel}
\usepackage[colorlinks]{hyperref}
\usepackage[usenames,dvipsnames]{color}
\hypersetup{
     breaklinks=true,
    pdfstartview={FitH},    
    colorlinks=true,       
    linkcolor=blue,          
    citecolor=red,        
    filecolor=magenta,      
    urlcolor=blue,           
    anchorcolor=green,      
    linktocpage=true
}

\def\doi{http://doi.org}

\begin{document}
\title{Observational constraints on the jerk parameter with the data of
the Hubble parameter}
\author{Abdulla Al Mamon}
\email{abdulla.physics@gmail.com}
\affiliation{Department of Mathematics, Jadavpur University, Kolkata-700032, India}
\author{Kazuharu Bamba}
\email{bamba@sss.fukushima-u.ac.jp}
\affiliation{Division of Human Support System, Faculty of Symbiotic
Systems Science, Fukushima University, Fukushima 960-1296, Japan}
\newcommand{\be}{\begin{equation}}
\newcommand{\ee}{\end{equation}}
\newcommand{\bea}{\begin{eqnarray}}
\newcommand{\eea}{\end{eqnarray}}
\newcommand{\bc}{\begin{center}}
\newcommand{\ec}{\end{center}}
\begin{abstract}
We study the accelerated expansion phase of the universe by using the {\textit{kinematic approach}}. In particular, the deceleration parameter $q$ is parametrized in a model-independent way. Considering a generalized parametrization for $q$, we first obtain the jerk parameter $j$ (a dimensionless third time derivative of the scale factor) and then confront it with cosmic observations. We use the latest observational dataset of the Hubble parameter $H(z)$ consisting of 41 data points in the redshift range of $0.07 \leq z \leq 2.36$, larger than the redshift range that covered by the Type Ia supernova. We also acquire the current values of the deceleration parameter $q_0$, jerk parameter $j_0$ and transition redshift $z_t$ (at which the expansion of the universe switches from being decelerated to accelerated) with $1\sigma$ errors ($68.3\%$ confidence level). As a result, it is demonstrate that the universe is indeed undergoing an accelerated expansion phase following the decelerated one. This is consistent with the present observations. Moreover, we find the departure for the present model from the standard $\Lambda$CDM model according to the evolution of $j$. Furthermore, the evolution of the normalized Hubble parameter is shown for the present model and it is compared with the dataset of $H(z)$. 
\end{abstract} 

\maketitle
\section{Introduction}
\par Since the end of last century, various independent observations \cite{riess1998,perl1999,teg2004,seljak2005,eisen2005,koma2011,hin2013,pl2014a,pl2015,pl2016a,pl2016b} have strongly suggested that the expansion of the universe is speeding up. But, understanding the fundamental physics behind this accelerated expansion is still an open question in modern cosmology. In order to give a reasonable explanation to this accelerating scenario, a large variety of attempts have been done. These attempts include the {\it modified gravity} models which relates to the changes of the geometry of the spacetime, and the {\it dark energy} models which involve the introduction of exotic matter sources (for a review, one can look into Refs. \cite{rev1,rev2,rev3,rev4,rev5,rev6,rev7,rev8}). In general, these models correspond to the {\it dynamics of the universe}. Although these models fit the observational data, but they also have their own demerits. For example, the $\Lambda$CDM model is the most natural one which shows very well consistence with the various observational data, however, it can not escape from the {\it fine tuning} and {\it cosmological coincidence}  problems \cite{ftp89,ccp99}. So, the study of explaining the cosmic acceleration is still continued.\\
\par In cosmology, another way to understand the cosmic acceleration is to analyze kinematic variables like the Hubble parameter ($H$), the deceleration parameter ($q$), or the jerk parameter ($j$), which are all derived from the derivatives of the scale factor (for details, see section \ref{sec2}). The {\it kinematic approach} is advantageous since it does not need any model specific assumptions like the composition of the universe. It is described by a metric theory of gravity and is assumed that the universe is homogeneous and isotropic at cosmological scales (for review on this topic, see \cite{km1,km2,km3,km4beg,km5beg,km6beg,jerk3,jerk4,turner2002,nair2012}). In the literature, there have been many attempts to constrain the present values of $H$, $q$ and $j$ by parametrizing $q$ or $j$ \cite{nair2012,kmp1,kmp2,kmp2beg,kmp3,kmp4,kmp5,kmp6,kmp7,kmp8,zhaij,tedebegell}. For example, Rapetti et al. \cite{jerk4} performed a systematic study of jerk parameter as the way towards building up a model in order to examine the expansion history of the universe.  On the other hand, by setting the $\Lambda$CDM model as the fiducial model and using the Type Ia Supernova and observational Hubble parameter data, recently Zhai et al. \cite{zhaij} constrained four jerk models  through different parametrizations of $j$ ($j=\Lambda$CDM value $+$ departure), as a function of the redshift $z$. In a pioneering work, Riess et al. \cite{qlin} measured a transition from an early decelerating to present accelerating phase using a simple linear redshift parameterization of $q$ ($q(z)=q_{0}+q_{1}z$). However, this model is not reliable at high redshift. In a recent work, Xu et al. \cite{kmp7} studied few kinematic models by considering the linear first order expansion of $q$ ($q(z)=q_{0}+\frac{q_{1}z}{1+z}$), constant jerk parameter and third order expansion of luminosity distance. Following the same line of thought, our main goal in this paper is to examine some simple kinematic model for the cosmic expansion based on a general parameterization for $q$. The features of this parametrization have been discussed in the next section. We have then derived the expression of $j(z)$ for this specific choice of $q$. However, our work is more general and also different from other similar works \cite{kmp7,qlin} in different ways. Firstly, similar to the work of Zhai et al. \cite{zhaij}, we do not assume here a flat $\Lambda$CDM model for the present universe a priori, but rather allow our model to behave in a more general way. Additionally, the present value of the jerk parameter is allowed to be fixed by the observational data. Secondly, in this work, we go one step further by studying the evolution of jerk parameter for a general deceleration parameter, which is independent of the matter content of the universe. Lastly, here, we employ the  latest $H(z)$ dataset as useful cosmic constraints.\\
\par The paper is organized as the following. In section \ref{sec2}, we have described the phenomenological model considered here. In section \ref{data}, we have described the observational data used in the present work along with the statistical analysis and discussed the results in section \ref{result}. Finally, we have summarized the main conclusions in section \ref{conclusion}.
\section{The Kinematic model}\label{sec2}
In what follows, we have assumed a homogeneous and spatially flat Friedmann-Robertson-Walker (FRW) metric described by the line element:
\be
ds^{2}=dt^{2} - a^{2}(t)[dr^{2}+ r^{2}d{\Omega}^{2}]
\ee
where $a(t)$ is the cosmic scale factor (which is scaled to be unity at
the current epoch, i.e., $a_{0}=1$) and $t$ is the cosmic time.\\
\par As discussed in the previous section, the Hubble parameter, deceleration parameter and jerk parameter are purely kinematical, since they are independent of any gravity theory, and all of them are only related to scale factor $a$ or redshift $z$ (since, $a=\frac{1}{1+z}$). In particular, the jerk parameter, a dimensionless third derivative of the scale factor $a(t)$ with respect to cosmic time $t$, can provide us the simplest approach to search for departures from the concordance $\Lambda$CDM model. It is defined as \cite{jerk0tc,jerk1,jerk2,jerk3,jerk4}
\be
j(a)=\frac{\frac{d^3a}{dt^3}}{aH^3} 
\ee
where $H=\frac{\dot{a}}{a}$ is the Hubble parameter and the ``dot" implies derivative with respect to $t$. In terms of the deceleration parameter $q$ (a dimensionless second derivative of $a(t)$ with respect to $t$), the jerk parameter $j$ can be written as
\be\label{eqjerk}
j(q)={\left[q(2q+1) + (1+z)\frac{dq}{dz}\right]} 
\ee
where, $q=-\frac{\ddot{a}}{aH^2}$. Blandford et al. \cite{jerk3} described how the jerk parameterization provides an alternative and a convenient method to describe cosmological models close to concordance $\Lambda$CDM model. A powerful feature of $j$ is that for the $\Lambda$CDM model $j= 1$ (constant) always. It should be noted here that Sahni et al. \cite{jerkvs,jerkua} drew attention to the importance of $j$ for discriminating different dark energy models, because any deviation from the value of $j=1$ (just as deviations from the {\it equation of state} parameter $\omega_{\Lambda}=-1$ do in more standard dynamical approaches) would favor a non-$\Lambda$CDM model. The simplicity of the jerk formalism thus enables us to constrain the departure from the $\Lambda$CDM value in an effective manner. Also, the equation (\ref{eqjerk}) is useful when the parametric form of the deceleration parameter $q(z)$ is given. In fact, a variety of $q$-paramertized models have been proposed in the literature (for details see \cite{qlin,qcpl1,qcpl2,turner2002,aksaru2014,chuna2008,chuna2009,del2012,nair2012,
mamon2016a,mamon2017b,aamqgen,santos2011,gong2006,gong2007,xu2008,xu2009}). Following this line of thought, in the present work, we are interested to investigate the evolution of $j$ for a general $q$-parametrized model given in \cite{aamqgen}. It is given by \\
\be\label{ans}
q(z)=q_{0}-q_{1}{\left[\frac{(1+z)^{-\alpha}-1}{\alpha}\right]}
\ee
where $q_{0}$, $q_{1}$ and $\alpha$ are arbitrary model parameters. In equation (\ref{ans}), $q_{0}$ indicates the present value of $q$, and $q_{1}$ indicates the derivative of $q(z)$ with respect to the redshift $z$.  From equation (\ref{ans}), one can easily recover few popular $q$-parametrized models in the following limits \cite{qlin,qcpl1,qcpl2}:
\bea\label{eqlimit}
{q(z)} = \left\{\begin{array}{ll} q_{0}+q_{1}z,&$for$\ \alpha= -1 \\\\
q_0 +q_{1}\ln(1+z),\ \ \ \ \ \ \ \ \ \ &$for$\
\alpha\rightarrow 0 \\\\
q_{0}+q_{1}{\left(\frac{z}{1+z}\right)},&$for$\ \alpha=+1
\end{array}\right..
\eea
It is worth noting here that the generalized parametrization is not valid at $\alpha=0$. The interesting cosmological characteristics of the parametrization, as given in equation (\ref{ans}), are extensively discussed in \cite{aamqgen}. Our main goal in this paper is to examine some simple kinematic model for the cosmic expansion based on the parameterization for $q(z)$ in equation (\ref{ans}). With this choice of $q(z)$, the expression for the Hubble parameter is obtained as
\be\label{eqhz}
H(z)= H_{0}{\rm exp}{\Big[\int^{z}_{0}\frac{1+q(x)}{1+x}dx\Big]}=H_{0}(1+z)^{(1+q_{0}+\frac{q_{1}}{\alpha})} {\rm exp}{\Big[\frac{q_{1}\left\{(1+z)^{-\alpha}-1\right\}}{\alpha^{2}}\Big]}
\ee
where $H_{0}$ denotes the present value of the Hubble parameter. In this case, the transition redshift (where ${\ddot{a}}(t)$ vanishes) can be obtained as
\be
z_{t}={\Big(\frac{q_{1}}{\alpha q_{0}+q_{1}}\Big)}^{\frac{1}{\alpha}}-1 
\ee
With the corresponding $q$-parametrization, the jerk parameter $j(z)$, defined in equation (\ref{eqjerk}), is obtained as
\be\label{eqjerkgen}
j(z)=q_{1}(1+z)^{-\alpha} + \alpha^{-2} (\alpha q_{0}+q_{1}-q_{1}(1+z)^{-\alpha})(\alpha +2\alpha q_{0}+2q_{1}-2q_{1}(1+z)^{-\alpha})
\ee
with $j_{0}=j(z=0)=q_{1}+q_{0}(1+2q_{0})$. Since the expression for the jerk parameter is explicit, so we can think that we are actually parameterizing $j(z)$ instead of $q(z)$. The advantage of this type of jerk parametrization is that it incorporates a wide class of viable models of cosmic evolution based on the choice of the parameter $\alpha$. Similar to the generalized $q$-parametrization, the above jerk parametrization is also not valid at $\alpha=0$. As mentioned in the previous section, Zhai et al. \cite{zhaij} parameterized $j(z)$ phenomenologically aiming at measuring the departure of $j$ from the $\Lambda$CDM value. An important difference with the work of Zhai et al. \cite{zhaij} is that we do not assume a flat $\Lambda$CDM model for the present universe a priori, but rather allow our model to behave in a more general way. Also, the value of $j_{0}$ is allowed to be fixed by the observational dataset.\\  
\par Obviously, the cosmological characteristics of the model given in equation (\ref{ans}) (or equation (\ref{eqjerkgen})) strongly depend on values of the parameters $q_{0}$, $q_{1}$ and $\alpha$. Using the latest $H(z)$ data, in the next section, we have constrained the parameters $q_{0}$ and $q_{1}$ for some specific values of $\alpha$.
\section{Observational constraints on the model parameters}\label{data}
In this section, we have described the latest observational data used in our analysis and the method used to analyze them.\\
\par It is known that the Type Ia Supernova or CMB or BAO dataset is powerful in constraining the cosmological models. However, the integration in its formula makes it hard to reflect the precise measurement of the expansion rate of the universe as a function of redshift, i.e., $H(z)$ \cite{sigmah}. This is the most direct and model independent observable of the dynamics of the universe. Therefore, the fine structure of the expansion history of the universe can be well indicated by the $H(z)$ dataset. From the observational point of view, the ages of the most massive and passively evolving galaxies will provide direct measurements of $H(z)$ at different redshifts, which develop another type of standard probe (namely, {\it standard clocks}) in cosmology \cite{jim2002}. It should be noted that $H(z)$ measurements are always obtained from two different techniques: galaxy differential age (also known as {\it cosmic chronometer}) and {\it radial BAO size} methods. The Hubble parameter depending on the differential ages as a function of redshift $z$ can be written in the form of
\be
H(z)=-\frac{1}{(1+z)}\frac{dz}{dt} 
\ee 
So, $H(z)$ can be obtained directly if $\frac{dz}{dt}$ is known \cite{simon2005}. Also, the apparently small uncertainty of this measurement naturally increases its weight in the $\chi^2$ statistics. For this dataset, the $\chi^{2}$ is defined as
\be\label{eqchi2h}
\chi^2_{H} = \sum^{41}_{i=1}\frac{[{H}^{obs}(z_{i}) - {H}^{th}(z_{i},H_{0},\theta)]^2}{\sigma^2_{H}(z_{i})}, 
\ee
where ${H}^{obs}$ is the observed Hubble parameter at $z_{i}$ and  ${ H}^{th}$ is the corresponding  theoretical value given by equation (\ref{eqhz}). Also, $\sigma_{H}(z_{i})$ represents the uncertainty for the $i^{th}$ data point in the sample and $\theta$ denotes the model parameter. In this work, we have used the latest observational $H(z)$ dataset consisting of 41 data points in the redshift range, $0.07\le z \le 2.36$, larger than the redshift range that covered by the Type Ia supernova. Among them, 5 new data points of H(z) are obtained from the differential age method by Moresco et al. \cite{hzdataMore} and 36 data points (10 data points are deduced from the radial BAO size method and 26 data points are obtained from the galaxy differential age method) are compiled by Meng et al. \cite{hzdataMeng}, Cao et al. \cite{hzdatarefcao} and Zhang et al. \cite{hzdatarefzhang}. In our analysis, we have also used the present value of Hubble parameter $H_0$, determined from the combined analysis with Planck+highL+WP+BAO \citep{pl2014b}. \\
\par The best fit values of the model parameters (say, $\theta^{*}$) from the Hubble data are estimated by minimizing the $\chi^2$ function in equation (\ref{eqchi2h}). It should be noted that the confidence levels $1\sigma(68.3\%)$ and $2\sigma(95.4\%)$ are taken proportional to $\bigtriangleup \chi^{2}=2.3$ and $6.17$ respectively, where $\bigtriangleup \chi^{2}= \chi^{2}(\theta)- \chi^{2}(\theta^{*})$ and $\chi^{2}_{m}$ is the minimum value of $\chi^2$. An important quantity which is used for data fitting process is
\be
{\bar{\chi}}^{2}=\frac{\chi^{2}_{m}}{dof} 
\ee
where subscript ``dof'' is the abbreviation of {\it degree of freedom}, and  it is defined as the difference between all observational data points and the number of free parameters. If $\frac{\chi^{2}_{m}}{dof} \le 1$, then the fit is good and the observed data are consistent with the considered model. 
\section{Results}\label{result} 
\begin{figure*}
\resizebox{5cm}{!}{\rotatebox{0}{\includegraphics{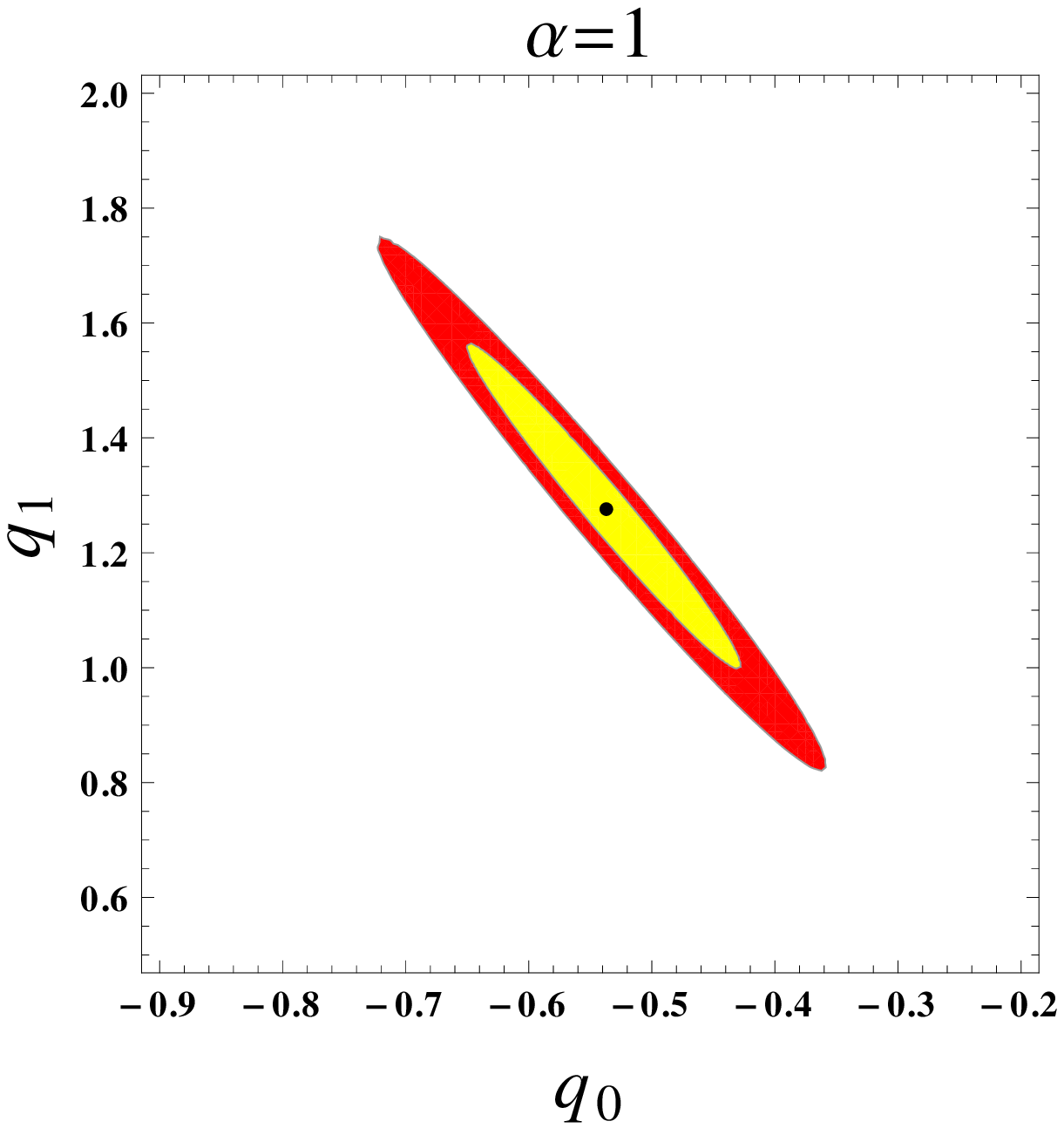}}}
\resizebox{5cm}{!}{\rotatebox{0}{\includegraphics{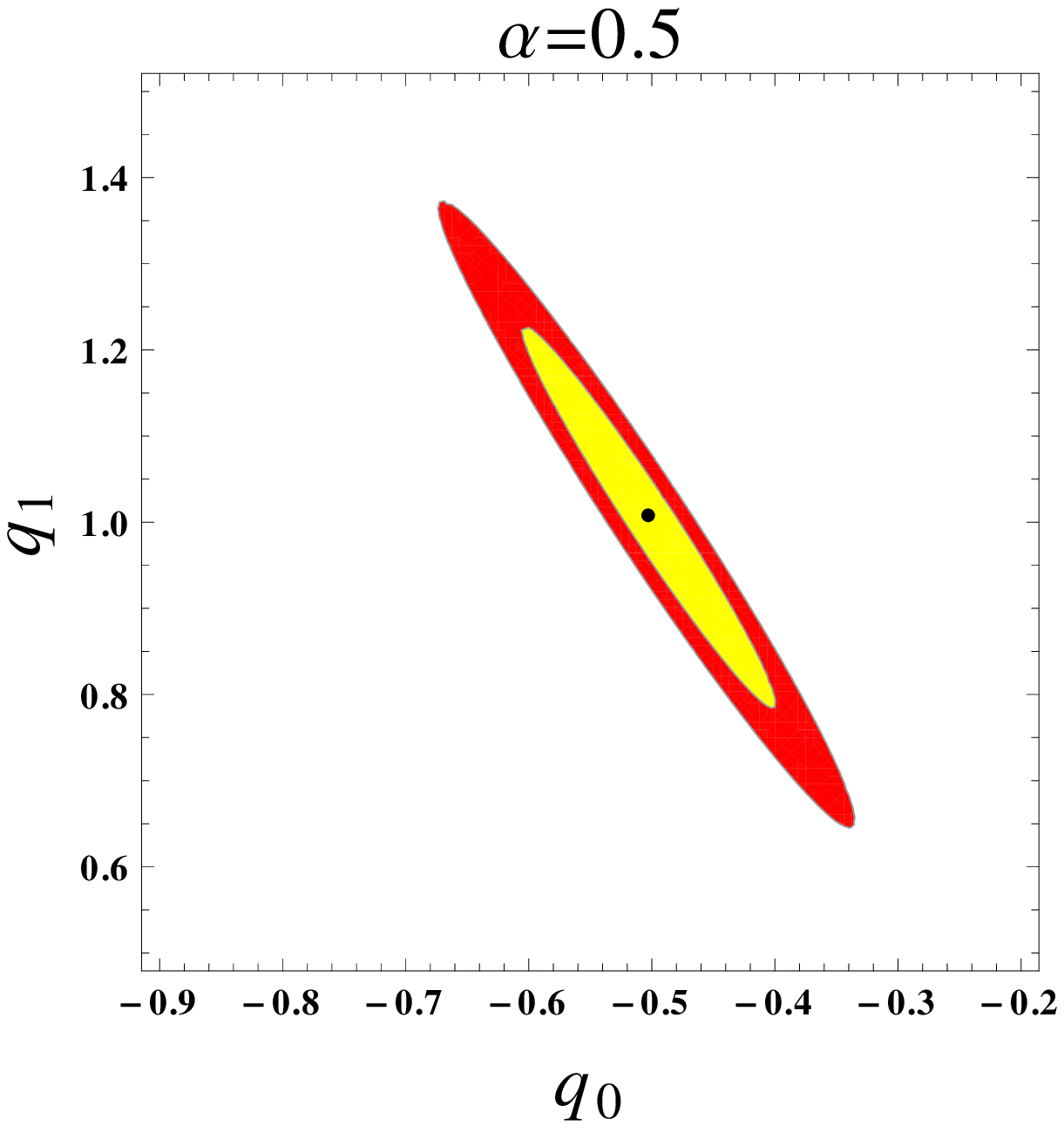}}}
\resizebox{5cm}{!}{\rotatebox{0}{\includegraphics{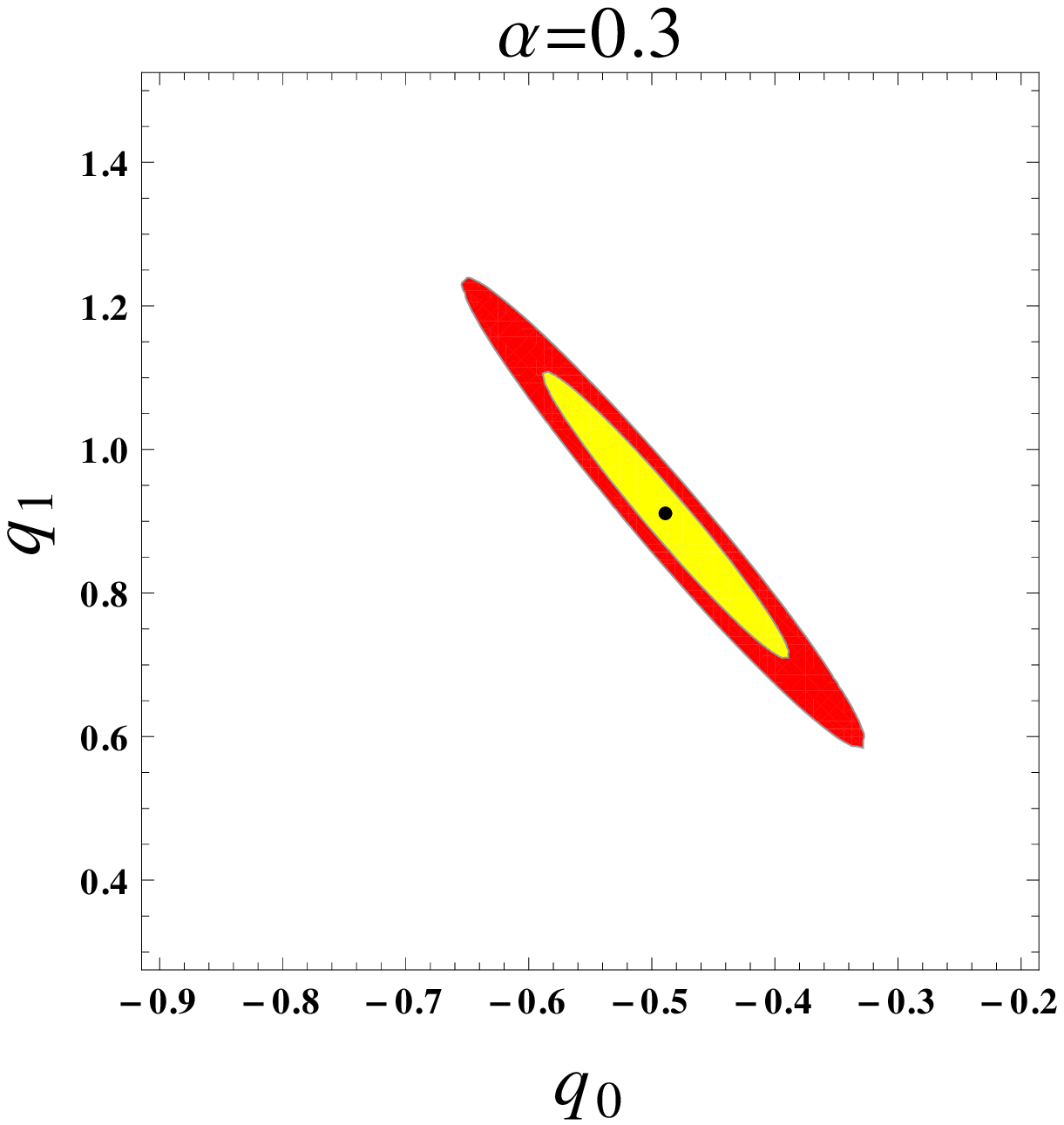}}}\\
\vspace{2mm}
\resizebox{5cm}{!}{\rotatebox{0}{\includegraphics{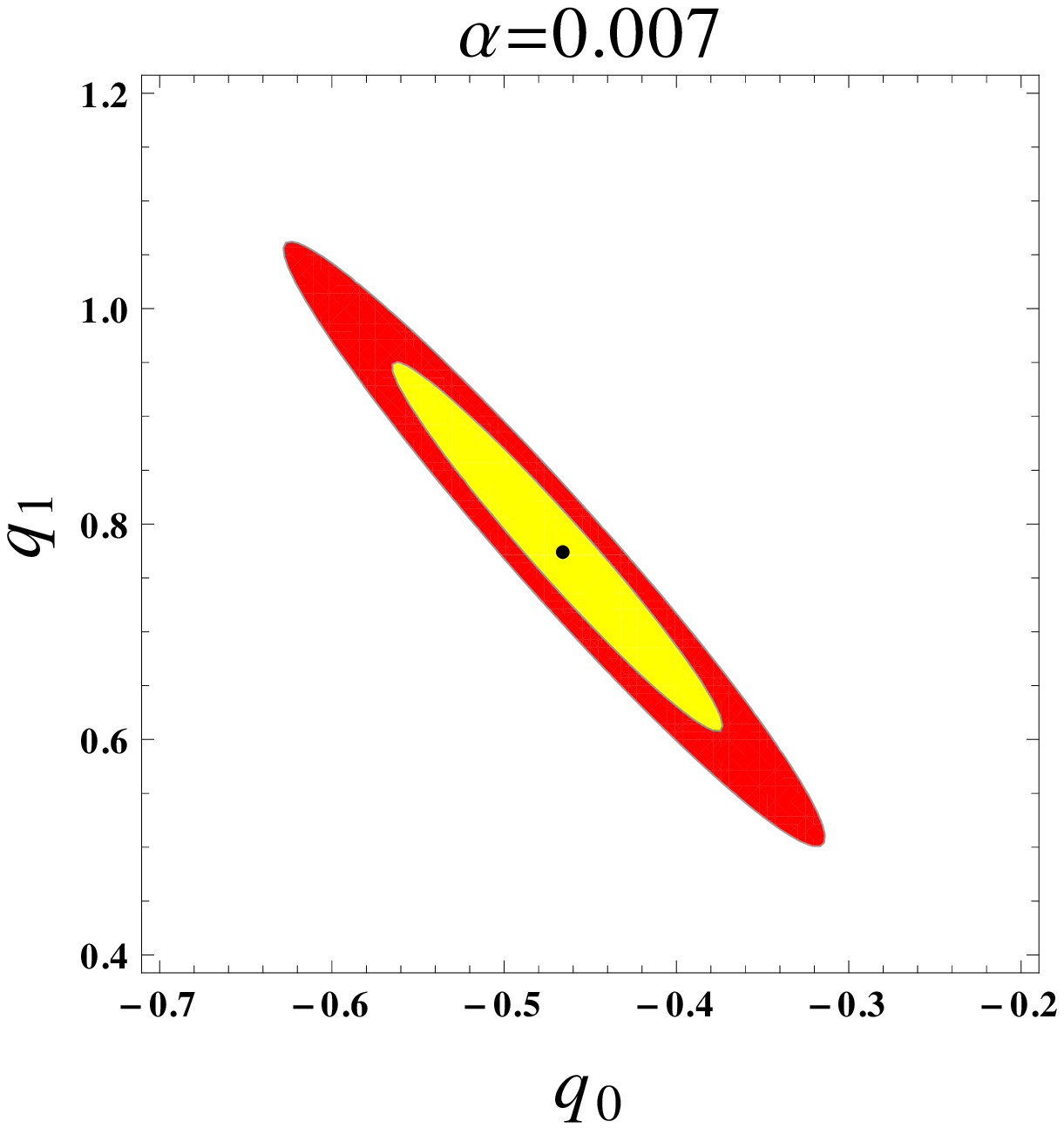}}}
\resizebox{5cm}{!}{\rotatebox{0}{\includegraphics{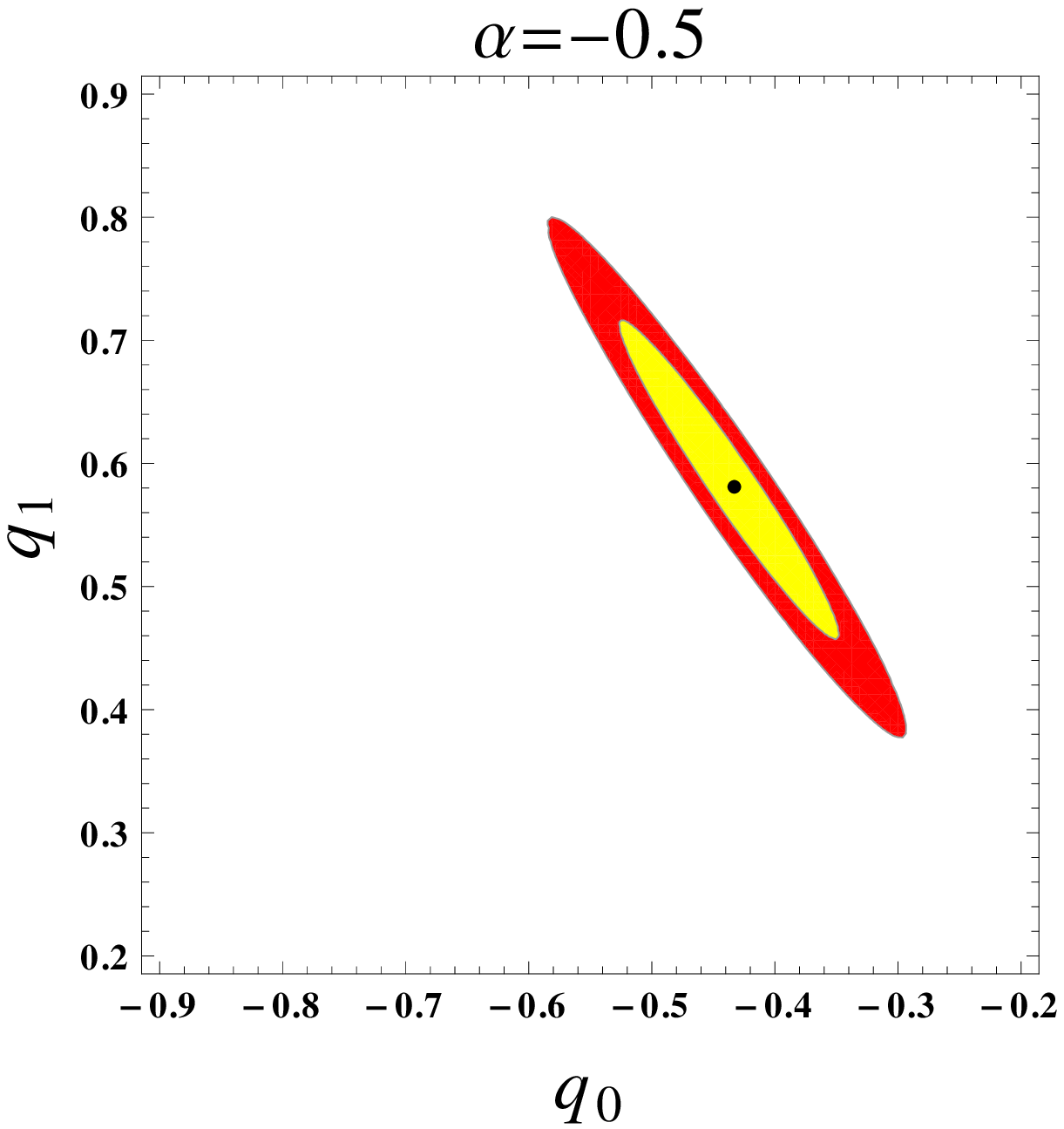}}}
\resizebox{5cm}{!}{\rotatebox{0}{\includegraphics{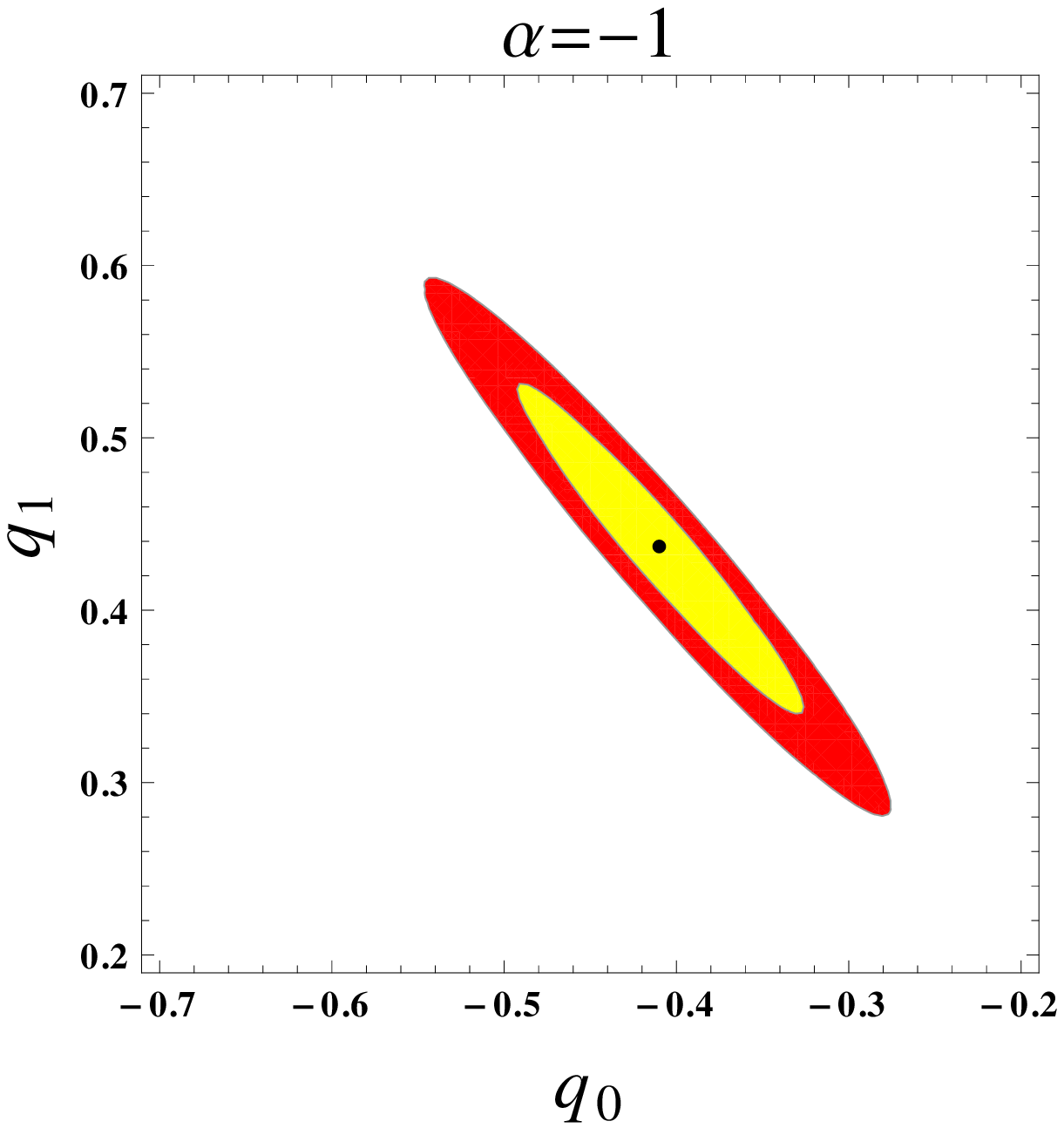}}}
\caption{This figure shows the $1\sigma$ (yellow region) and $2\sigma$ (red region) confidence contours in the $q_{0}$-$q_{1}$ plane using the latest $H(z)$ dataset and for different choices of $\alpha$, as indicated in each panel. In each panel, the black dot represents the best-fit values of the pair ($q_{0},q_{1}$).}
\label{figc}
\end{figure*}

Following the $\chi^{2}$ analysis (as described in the previous section), we have obtained the constraints on the model parameters $q_{0}$ and $q_{1}$  by fixing the parameter $\alpha$ to some constant values ($1$, $0.5$, $0.3$, $0.007$, $-0.5$ and $-1$) for the latest $H(z)$ dataset. The $1\sigma$ and $2\sigma$ contours in $q_{0}-q_{1}$ plane for the proposed model is shown in figure \ref{figc}. The best fit values with $1\sigma$ errors of the parameters $q_{0}$, $q_{1}$, $z_{t}$ and $j_{0}$ are displayed in table \ref{table1}. It is clear from table \ref{table1} that the model has almost same goodness in the viewpoint of ${\bar{\chi}}^2$ for different values of $\alpha$. But, there the values of $q_{0}$, $z_{t}$ and $j_{0}$ are different. \\
\par
\begin{figure*}
\resizebox{5cm}{!}{\rotatebox{0}{\includegraphics{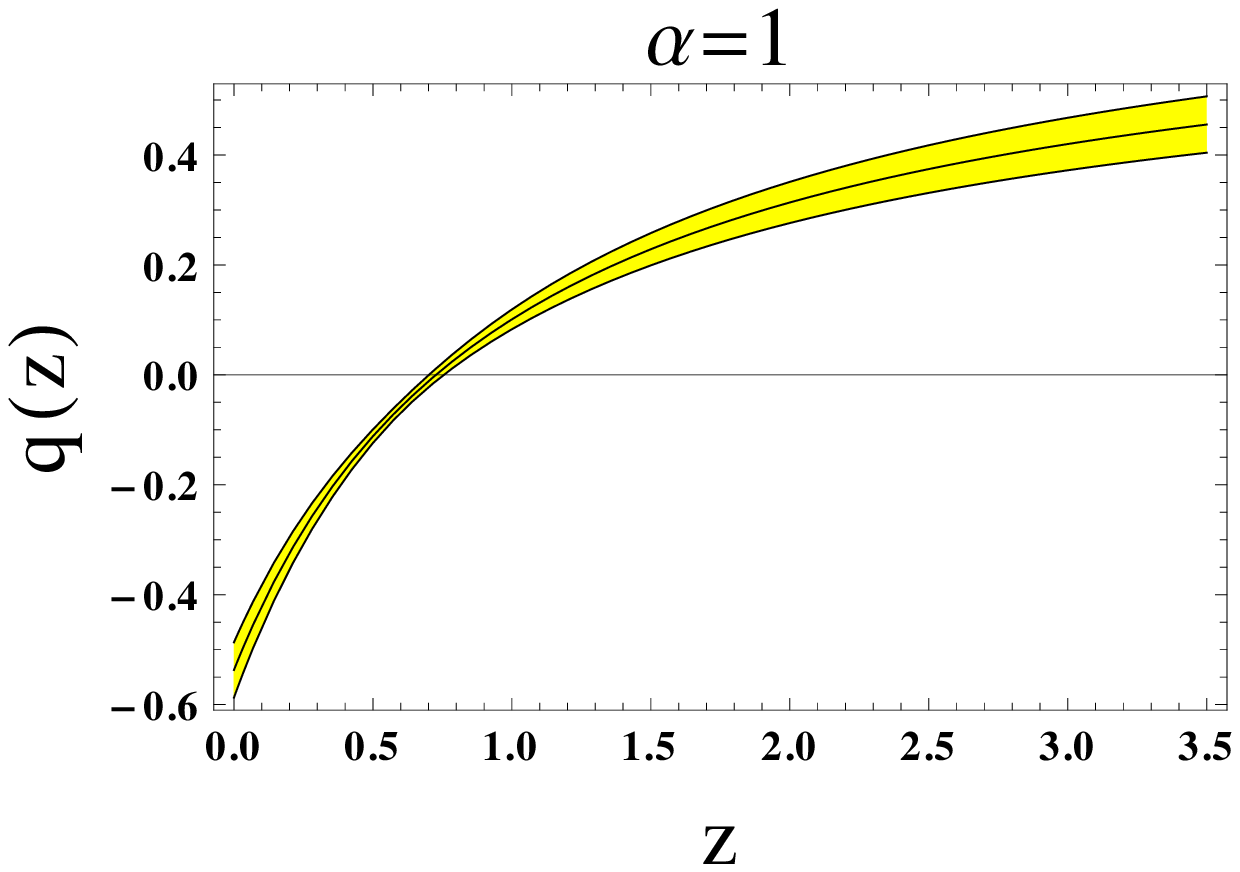}}}
\resizebox{5cm}{!}{\rotatebox{0}{\includegraphics{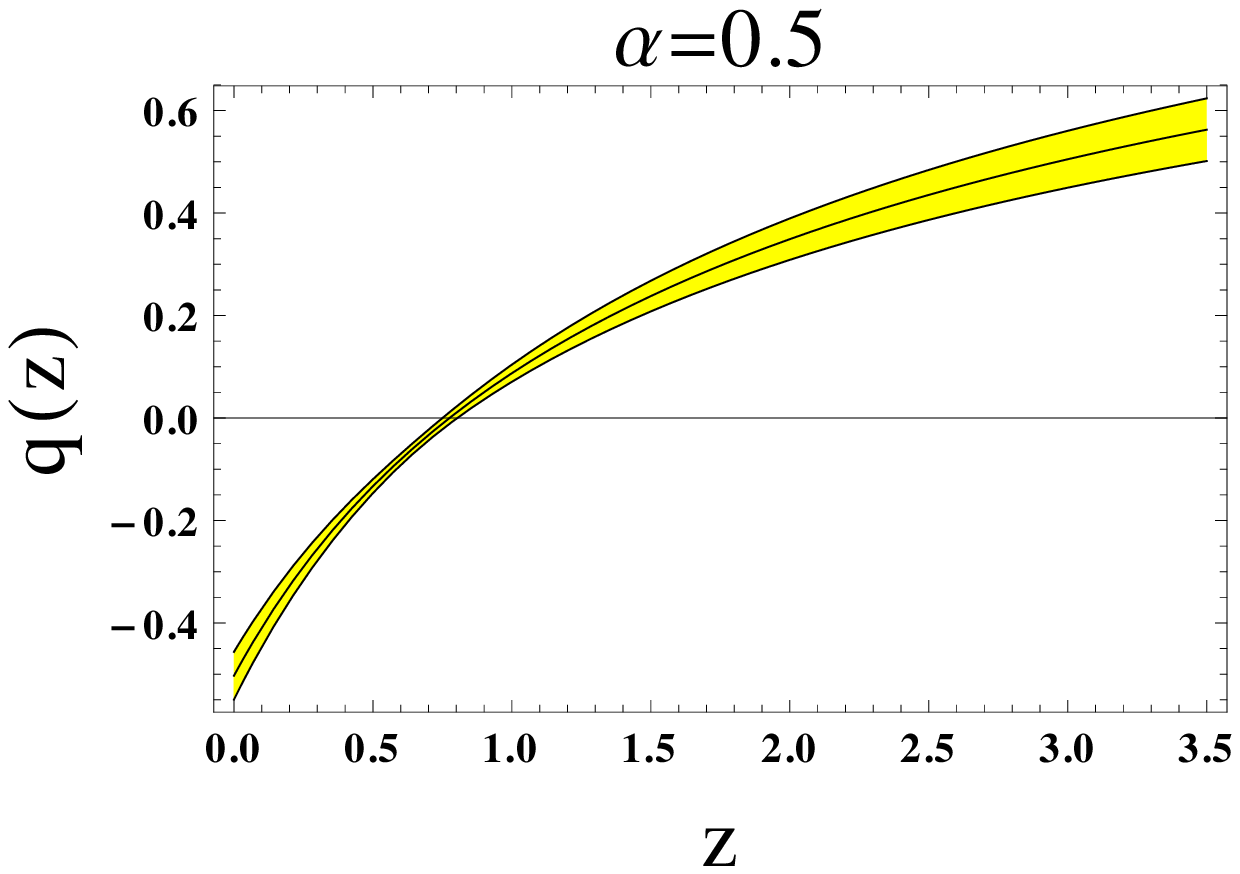}}}
\resizebox{5cm}{!}{\rotatebox{0}{\includegraphics{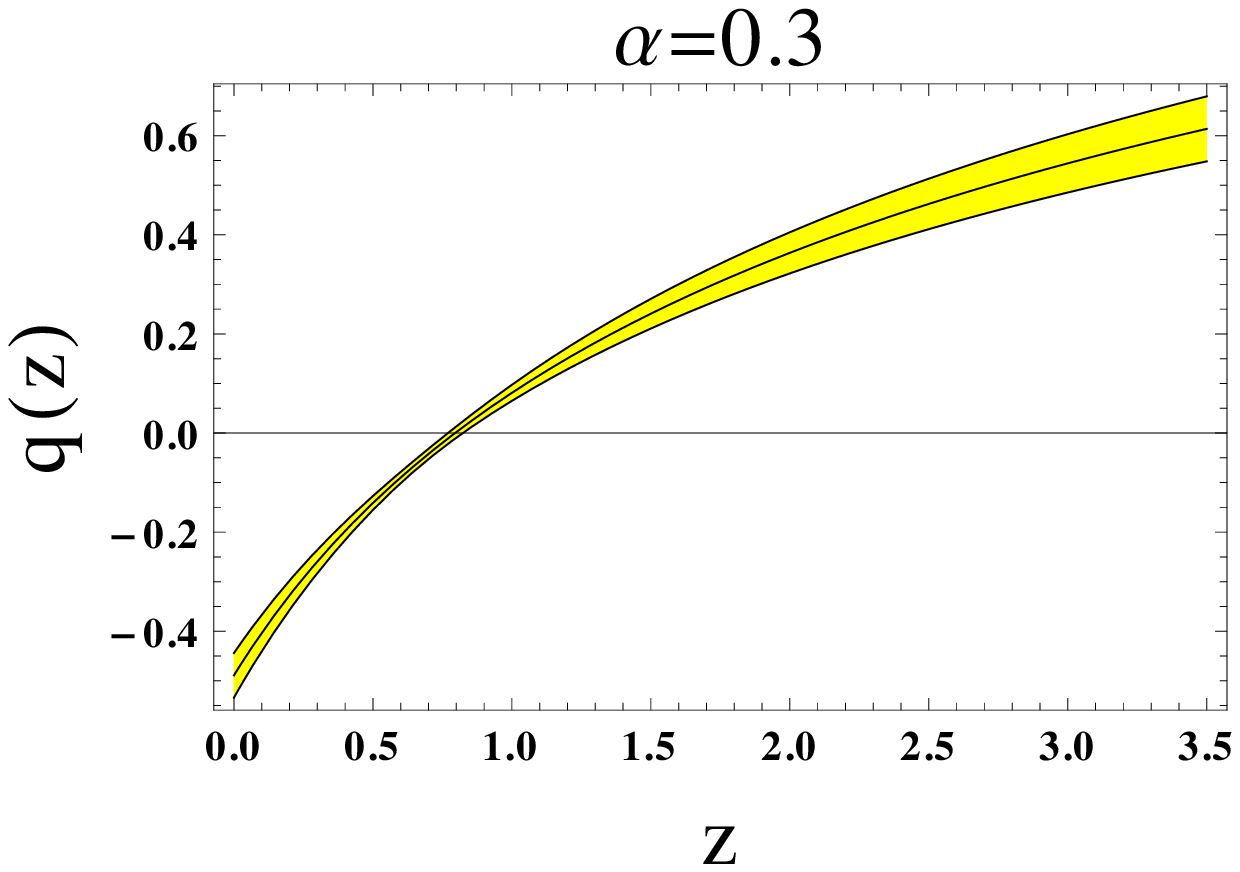}}}\\
\vspace{2mm}
\resizebox{5cm}{!}{\rotatebox{0}{\includegraphics{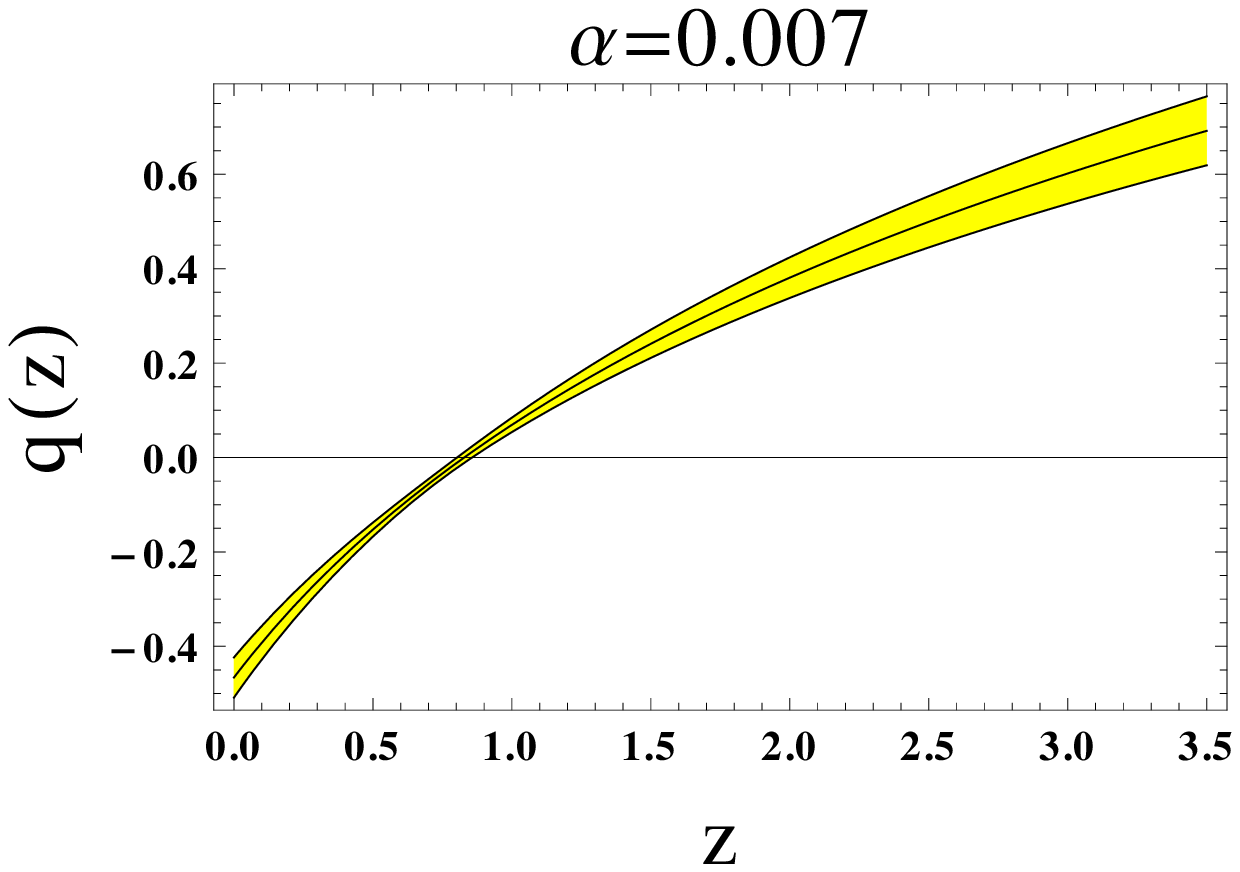}}}
\resizebox{5cm}{!}{\rotatebox{0}{\includegraphics{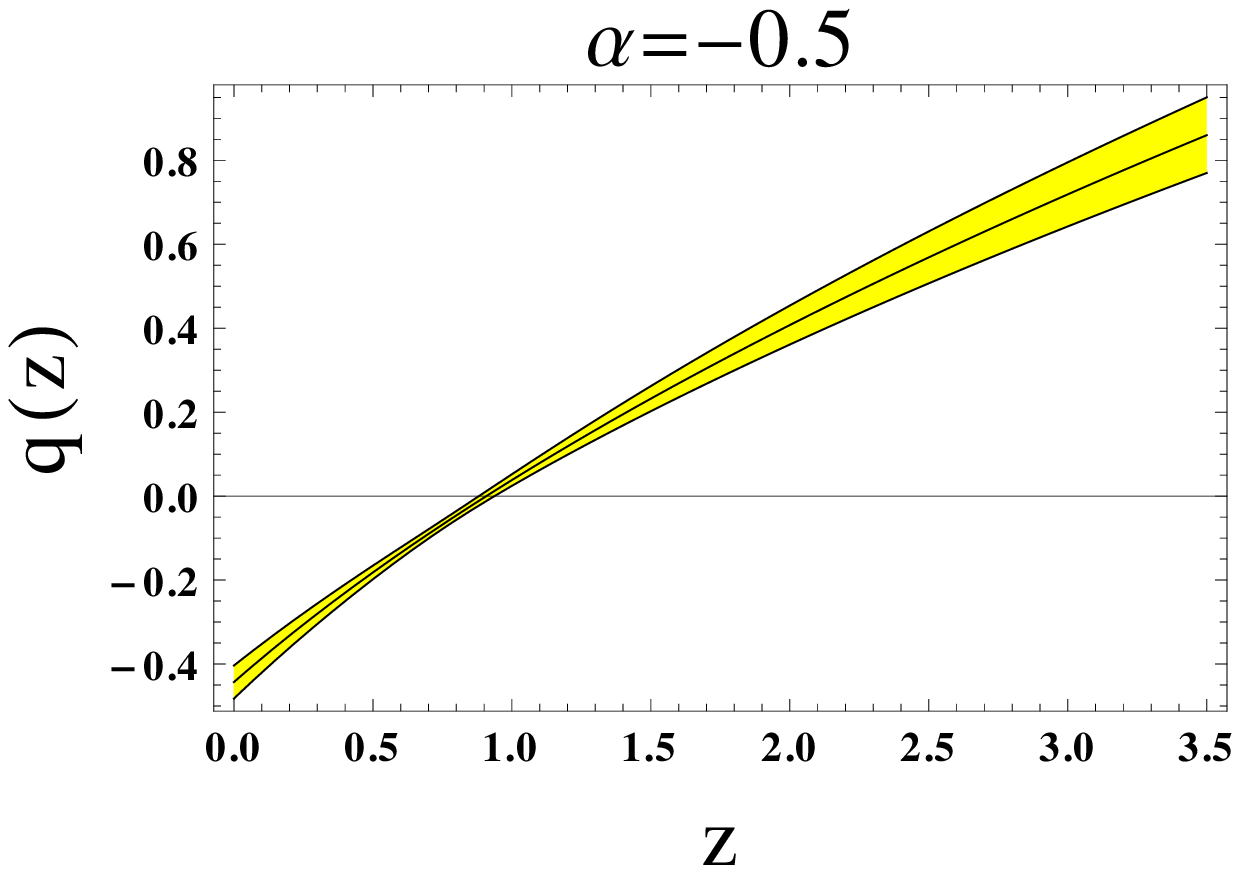}}}
\resizebox{5cm}{!}{\rotatebox{0}{\includegraphics{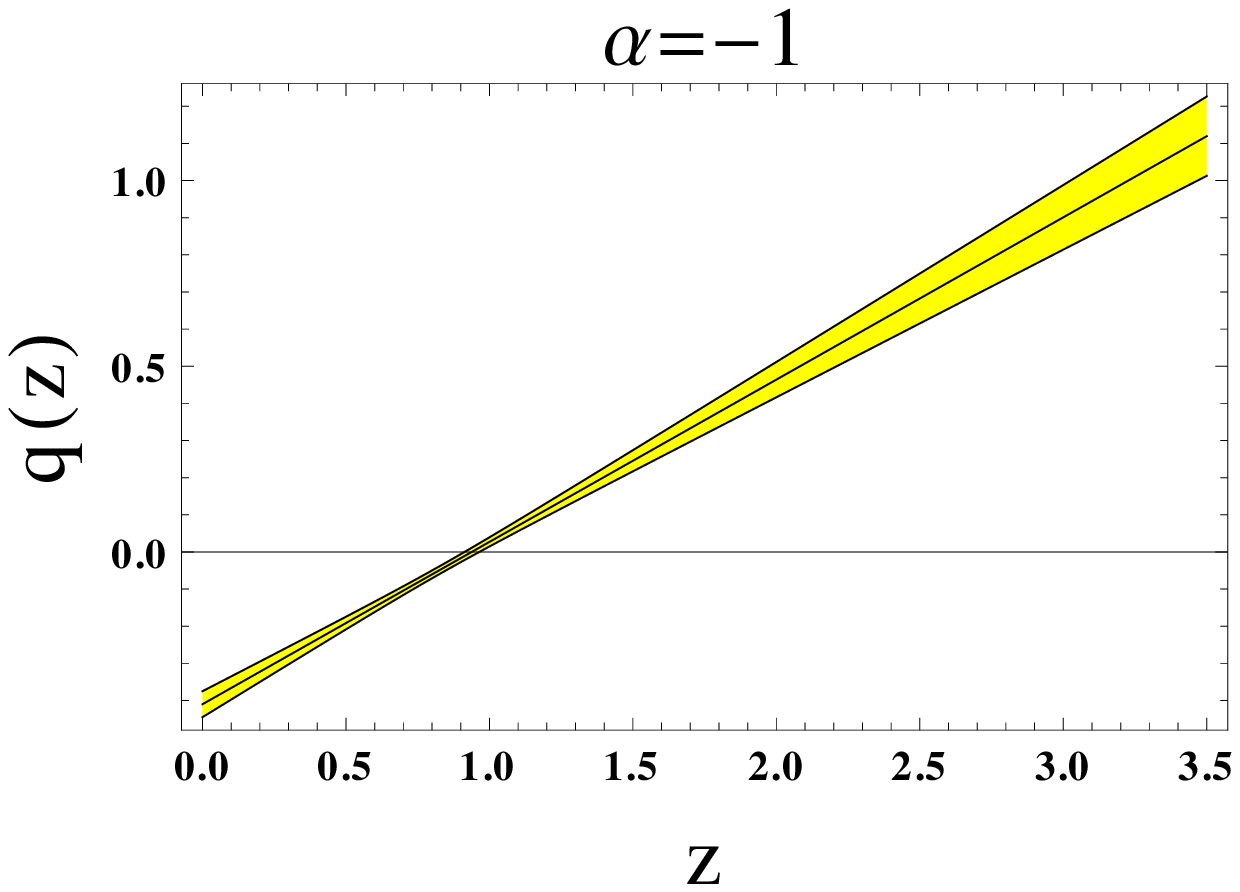}}}
\caption{Plots of the deceleration parameter $q(z)$ as a function of redshift $z$ are shown in $1\sigma$ error regions by considering different values of $\alpha$. In each panel, the central dark line denotes the best fit curve, while the horizontal line denotes $q(z)=0$.}
\label{figq}
\end{figure*}

 The plot of the deceleration parameter $q(z)$, as given in figure \ref{figq}, clearly shows that our model successfully generates late time cosmic acceleration along with a decelerated expansion in the
past for all values of $\alpha$. Figure \ref{figq} also shows that the transition from decelerated to accelerated expansion phase took place in the redshift range $0.685\le z \le 0.974$ within $1\sigma$ errors for different values of $\alpha$ (for details, see fifth column of table \ref{table1}). This range is well consistent with those previous results given in   \cite{jerk4,fr2013,capo2014,maga2014,nair2012,mamon2016a,mamon2017a,mamon2017b,fr2017}, though the transition redshift $z_t$ slightly increases as the value of $\alpha$ decreases.\\
\par
\begin{figure*}
\resizebox{5cm}{!}{\rotatebox{0}{\includegraphics{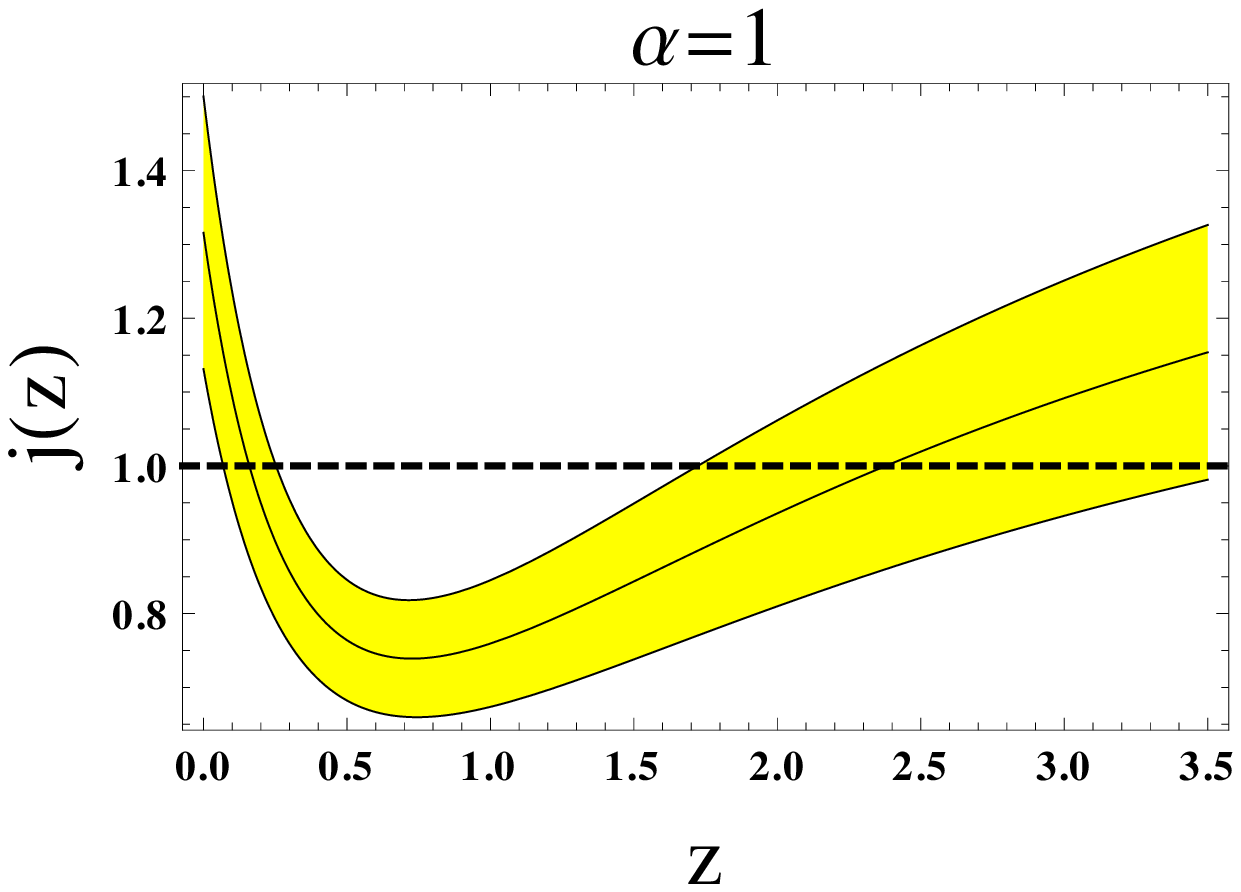}}}
\resizebox{5cm}{!}{\rotatebox{0}{\includegraphics{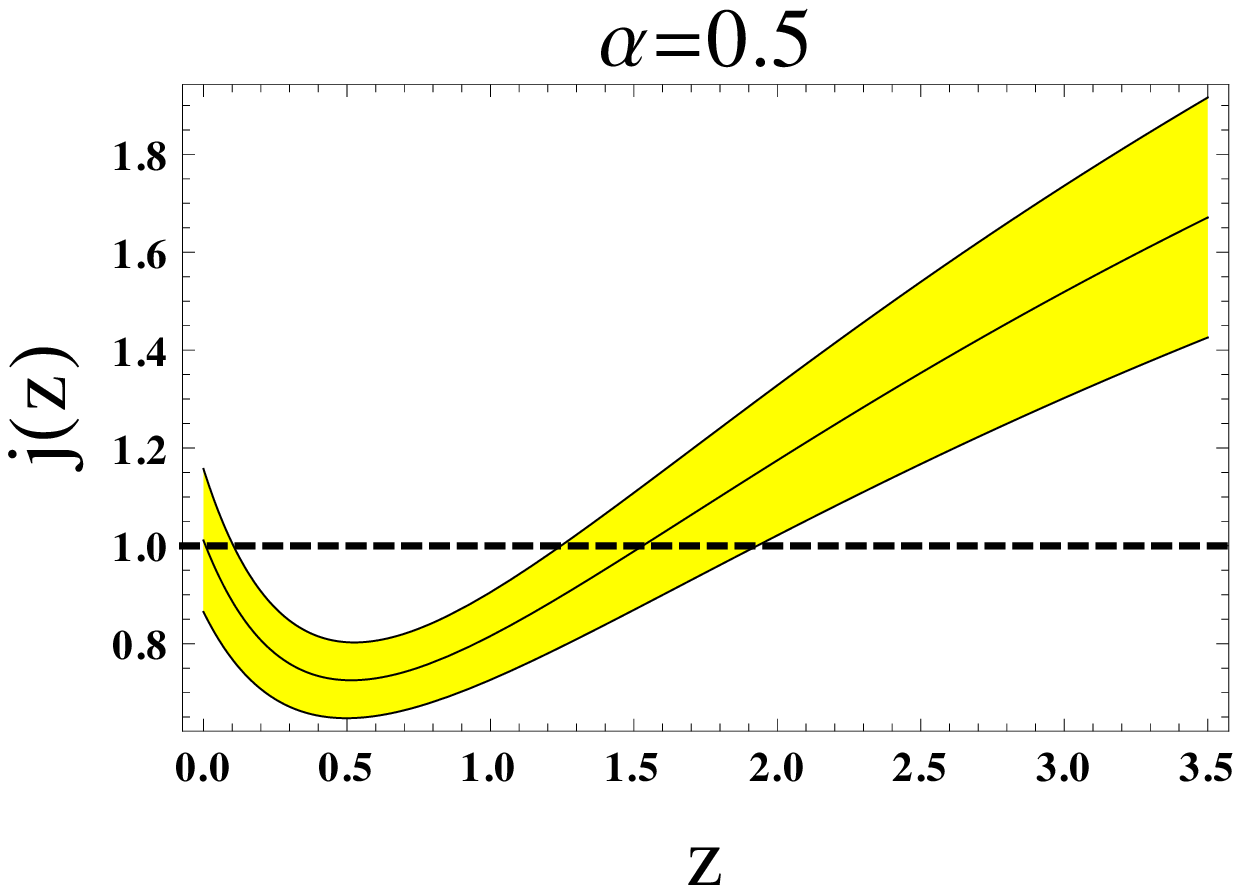}}}
\resizebox{5cm}{!}{\rotatebox{0}{\includegraphics{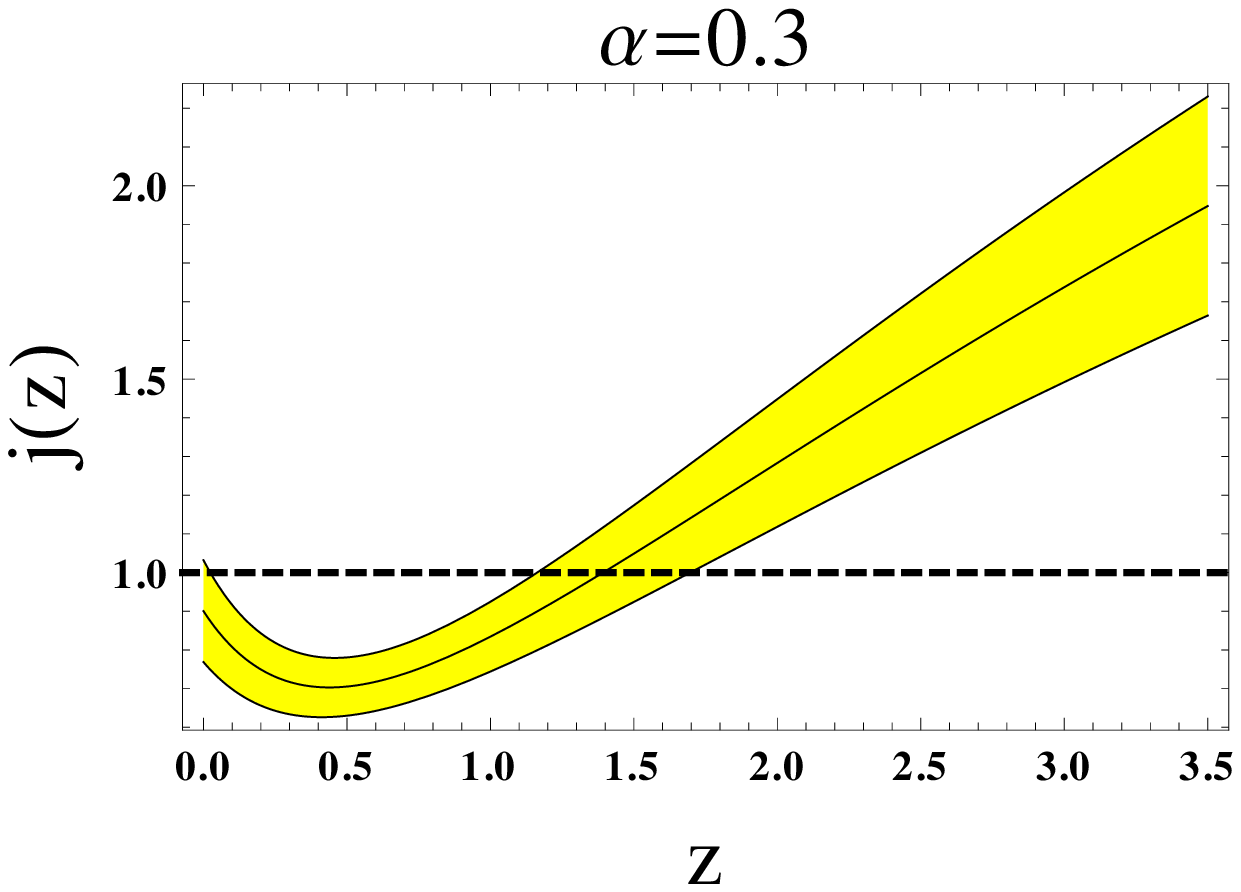}}}\\
\vspace{2mm}
\resizebox{5cm}{!}{\rotatebox{0}{\includegraphics{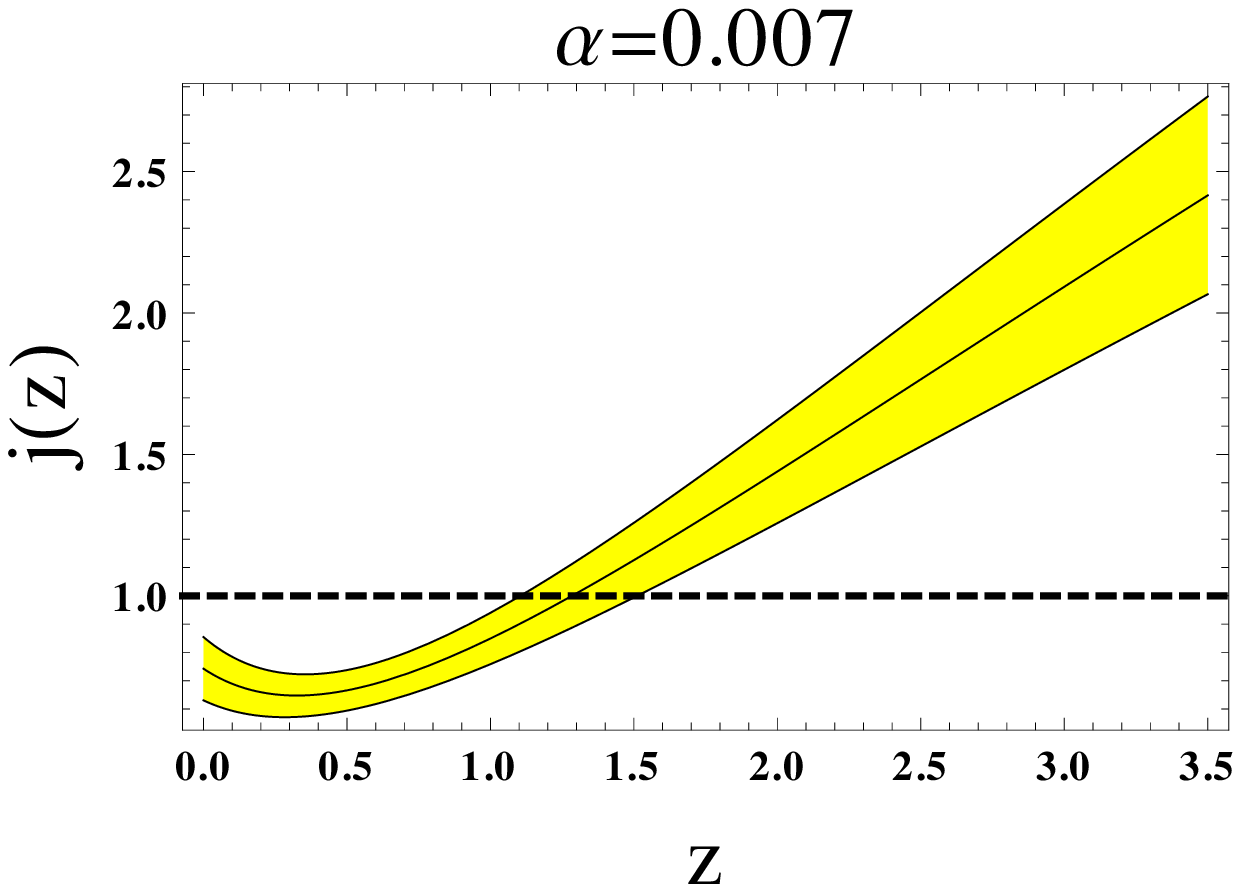}}}
\resizebox{5cm}{!}{\rotatebox{0}{\includegraphics{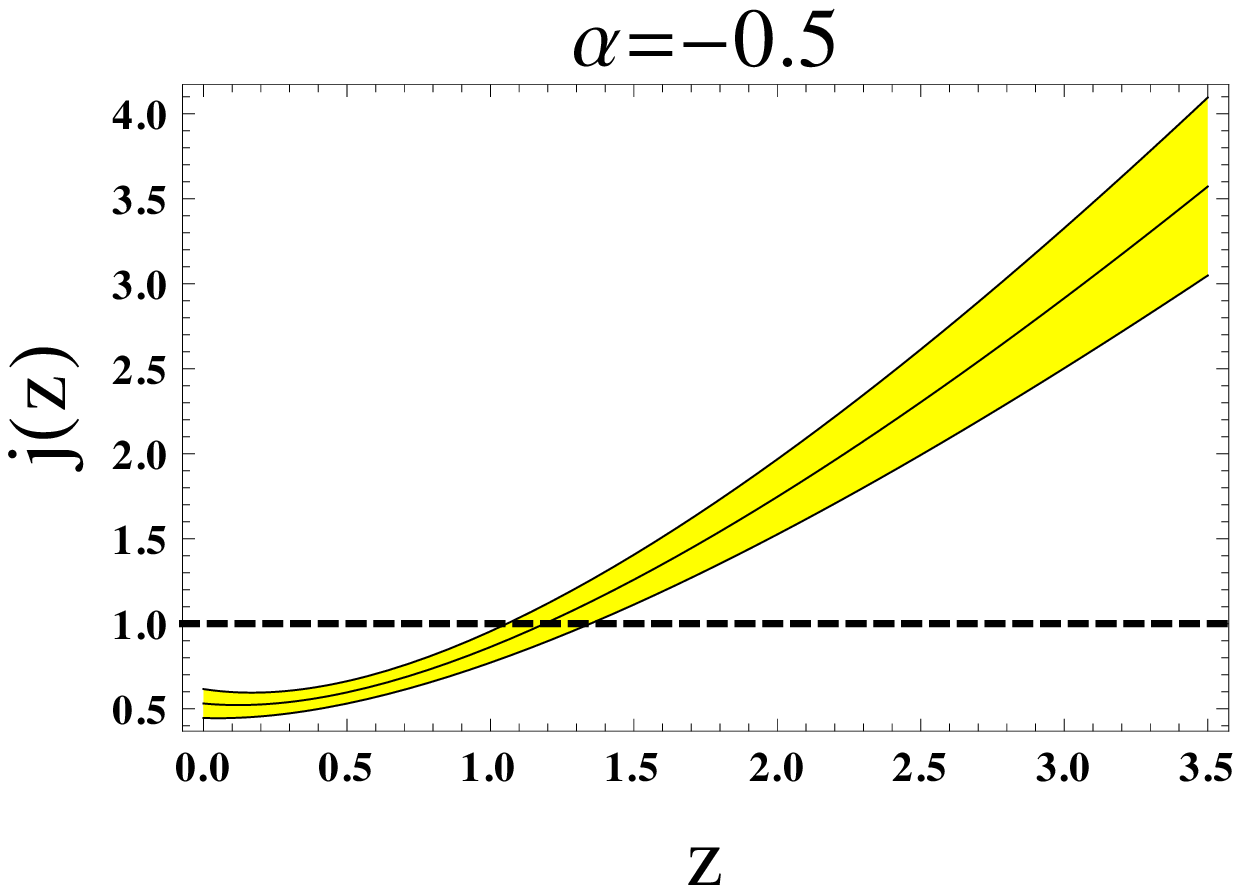}}}
\resizebox{5cm}{!}{\rotatebox{0}{\includegraphics{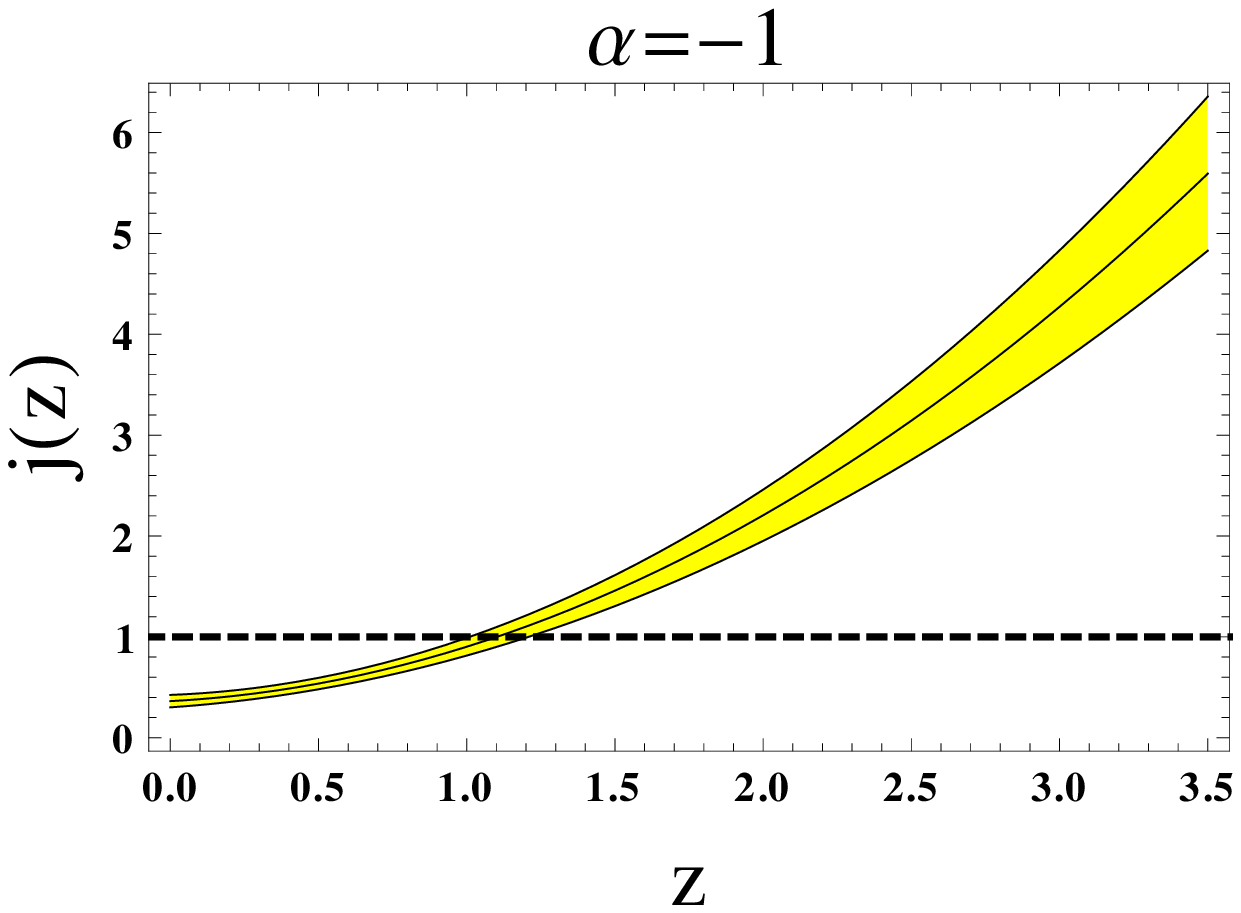}}}
\caption{Plots of the jerk parameter $j(z)$ as a function of redshift $z$ are shown in $1\sigma$ error regions by considering different values of $\alpha$. In each panel, the central dark line denotes the best fit curve, while the horizontal dashed line represents the concordance $\Lambda$CDM ($j=1$) model.}
\label{figj}
\end{figure*}

 Similarly, the evolutions of the jerk parameter $j(z)$ within $1\sigma$ error regions are shown in figure \ref{figj} for different values of $\alpha$. It has been found from figure \ref{figj} that the concordance $\Lambda$CDM model (dashed line) is not compatible within $1\sigma$ confidence level at the present epoch (i.e., $z=0$) for $\alpha=1,$ $0.007$, $-0.5$ and $-1$. It has also been found that the model (for $\alpha=1,$ $0.007$, $-0.5$ and $-1$) does not deviate very far from the $\Lambda$CDM model at the current epoch. These deviations of $j_{0}$ from the $\Lambda$CDM value also need attention as the real cause behind the cosmic acceleration is still unknown. However, it is also observed from figure \ref{figj} that for $\alpha=0.3$, the $\Lambda$CDM model is just marginally consistent within $1\sigma$ confidence level at $z=0$. On the other hand, the $\Lambda$CDM model is well consistent within $1\sigma$ confidence level at the present epoch for $\alpha=0.5$. Hence, our results (for $\alpha=0.5$ and $0.3$) also incorporate the flat $\Lambda$CDM model well within the $1\sigma$ error region like the work on the reconstruction of $j(z)$ by Zhai et al. \cite{zhaij}. The main difference is that Zhai et al. \cite{zhaij} imposed the jerk parameter to mimic as the flat $\Lambda$CDM model via their parametrization at $z=0$, but our work relaxes that requirement. In a word, the constraint results of $j_{0}$ tend to favor the dynamical jerk parameter. \\
\par
\begin{figure*}
\resizebox{5cm}{!}{\rotatebox{0}{\includegraphics{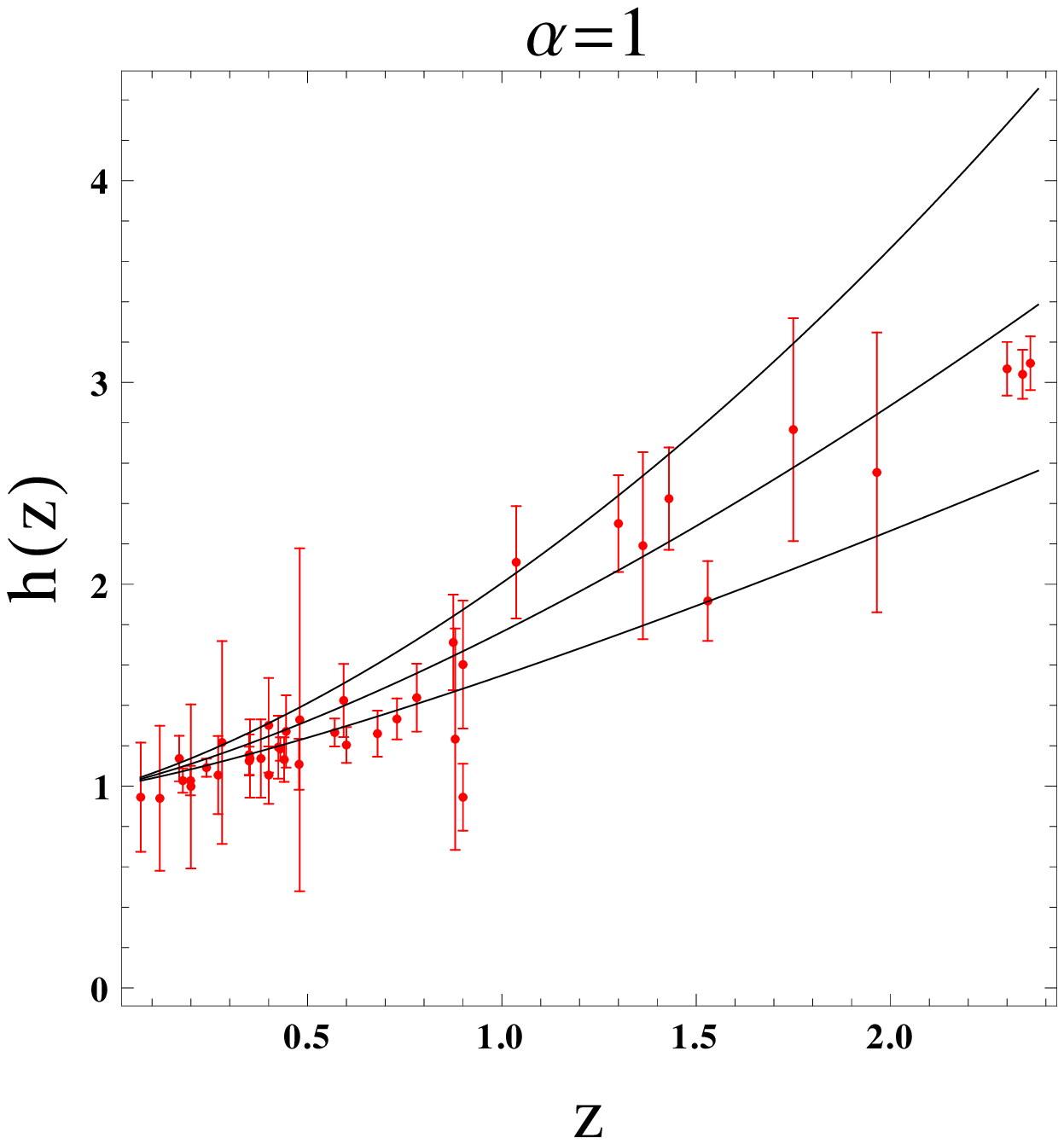}}}
\resizebox{5cm}{!}{\rotatebox{0}{\includegraphics{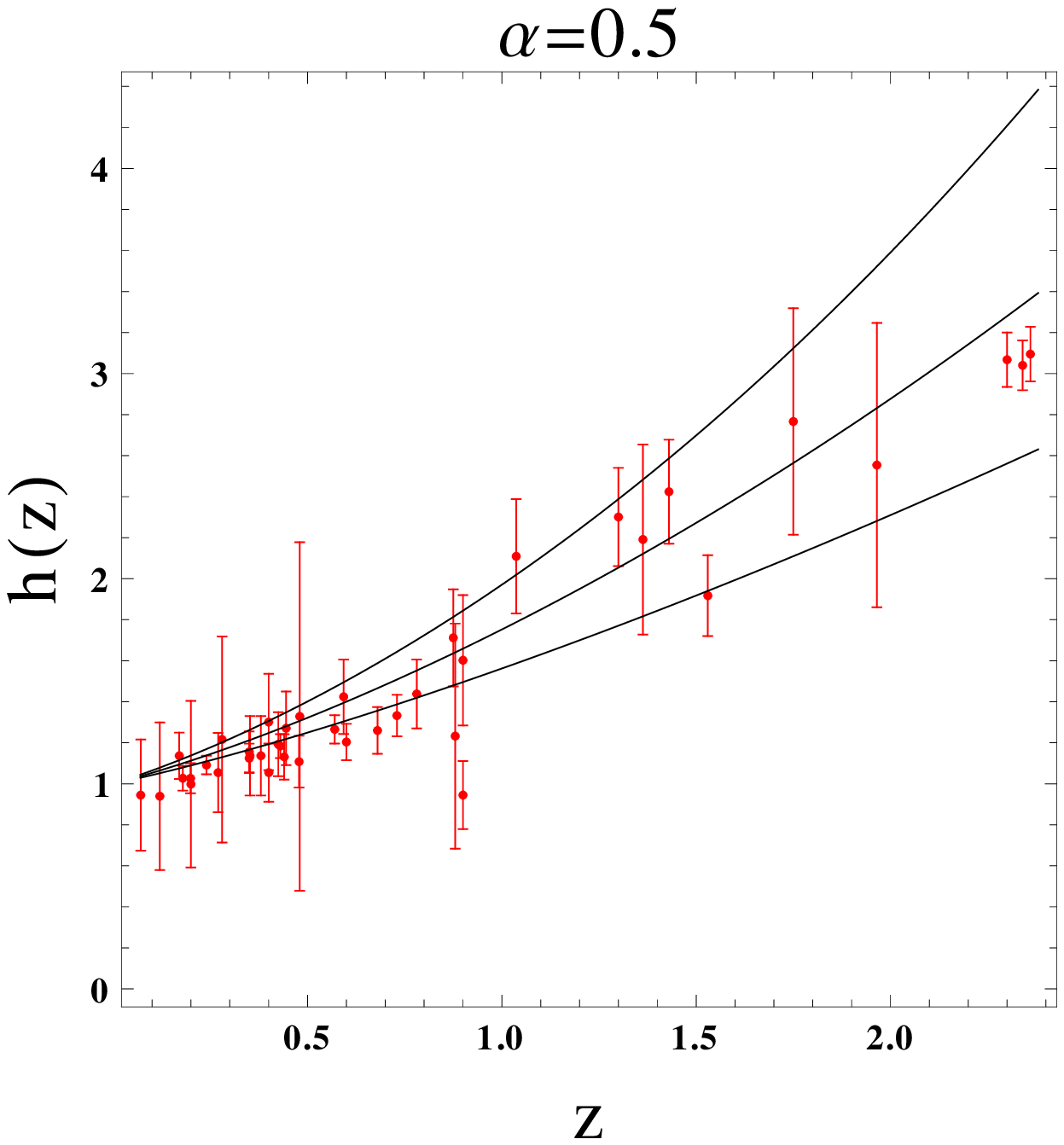}}}
\resizebox{5cm}{!}{\rotatebox{0}{\includegraphics{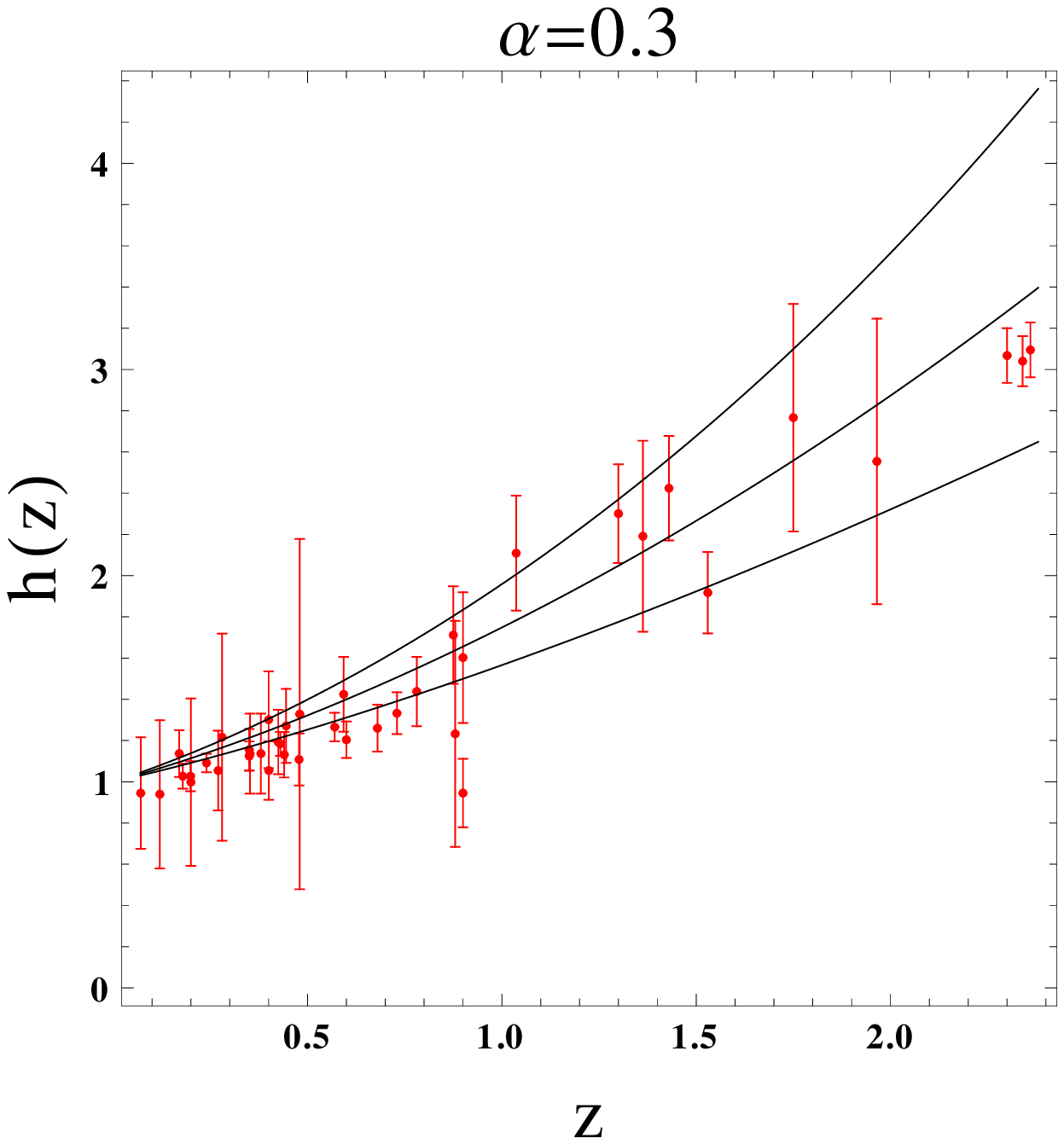}}}\\
\vspace{2mm}
\resizebox{5cm}{!}{\rotatebox{0}{\includegraphics{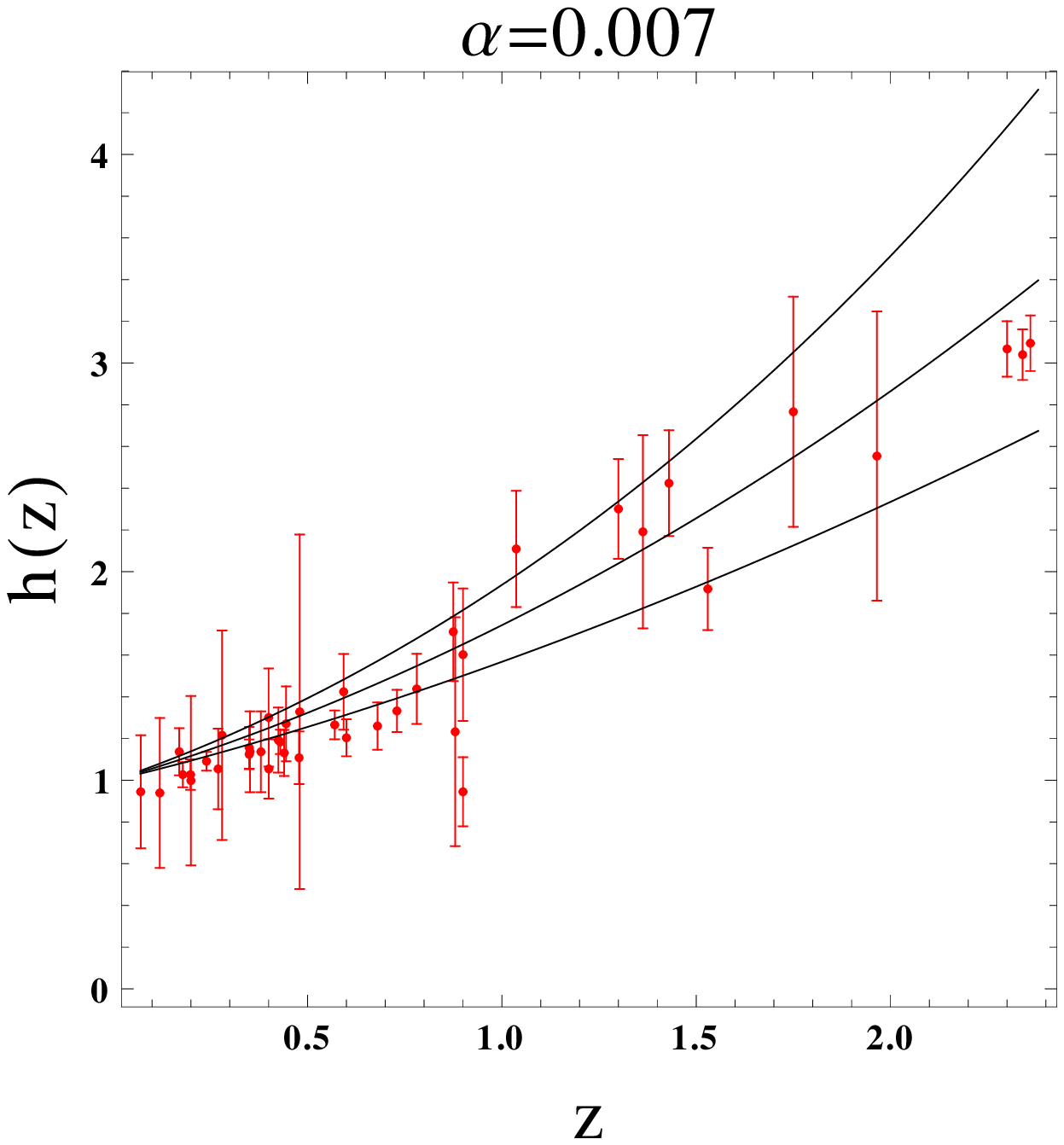}}}
\resizebox{5cm}{!}{\rotatebox{0}{\includegraphics{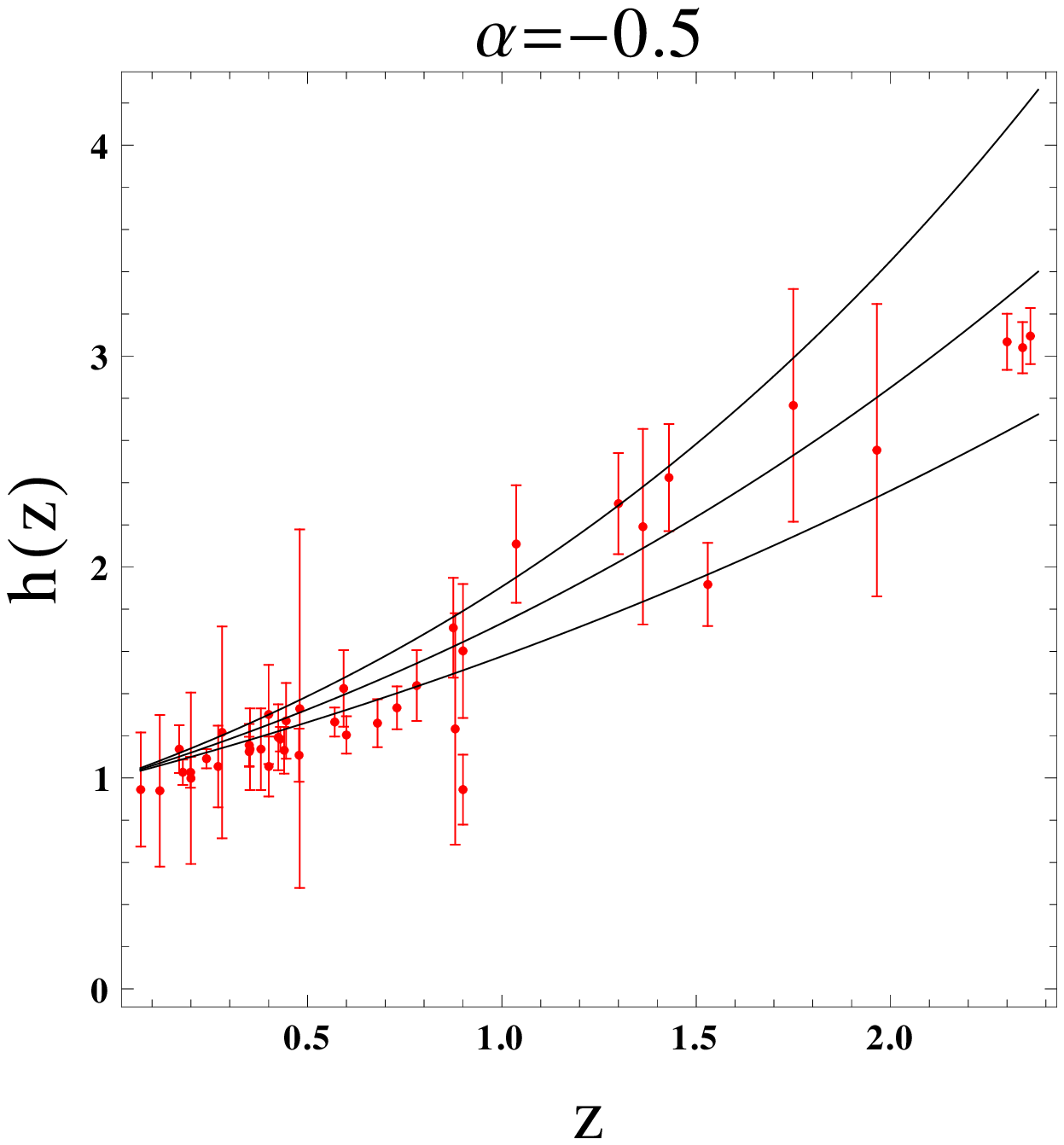}}}
\resizebox{5cm}{!}{\rotatebox{0}{\includegraphics{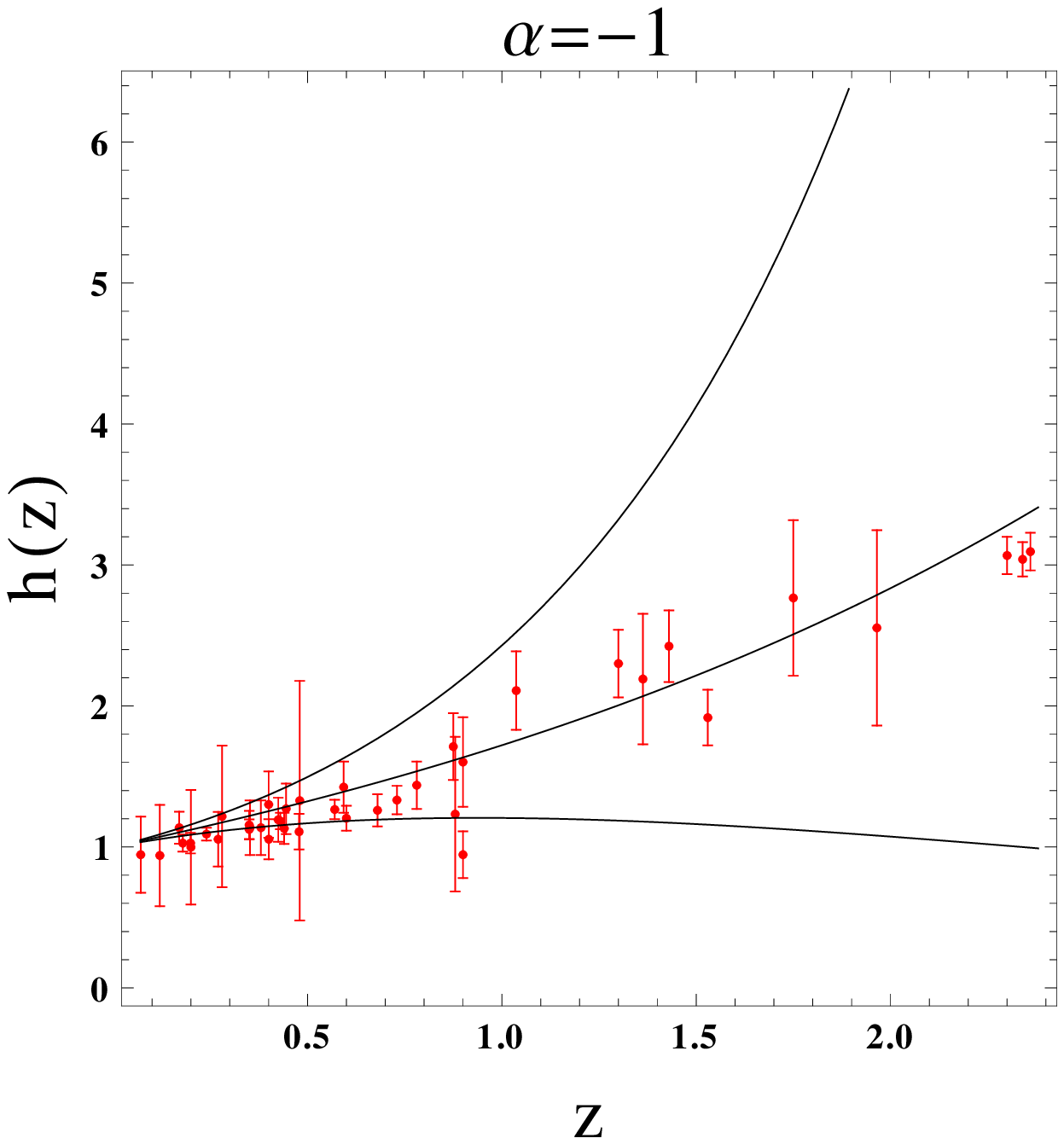}}}
\caption{Figure shows the evolution of the normalized Hubble parameter $h(z)=\frac{H(z)}{H_{0}}$, as a function of $z$, for the present model. We have also plotted $h(z)$ data (red dots) with $1\sigma$ error bars calculated from the compilation of 41 points $H(z)$ data \cite{hzdataMore,hzdataMeng}. In each panel, the central dark line denotes the best fit curve.}
\label{figh}
\end{figure*}

 In figure \ref{figh}, we have shown the evolution of the normalized Hubble parameter $h(z)=\frac{H(z)}{H_{0}}$ for our model and have compared that with the latest 41 points of $H(z)$ dataset \cite{hzdataMore,hzdataMeng}.  We have also plotted data points for $h(z)$ with $1\sigma$ error bars which have been obtained from the $H(z)$ dataset  using the current value of $H(z)$ given by Planck observations \citep{pl2014b}. The corresponding error in $h$ can be estimated as \cite{mamon2016a}
\be
\sigma_{h}=h \sqrt{\frac{\sigma^2_{H_{0}}}{H^2_0} + \frac{\sigma^2_{H}}{H^2}} 
\ee
where, $\sigma_{H_0}$ and $\sigma_{H}$ are the errors in $H_0$ and $H$ measurements respectively. We have observed from figure \ref{figh} that the our model is well consistent with the $H(z)$ data against redshift parameter for different values of $\alpha$ (except $\alpha =-1$ case). The reason is simple that the model with $\alpha =-1$ (i.e., $q(z)=q_{0}+q_{1}z$, see equation (\ref{eqlimit})) is not reliable at high redshift.
\begin{table*}
\caption{Best fit values of $q_{0}$ and $q_{1}$ with $1\sigma$ error bars obtained via $\chi^2$ minimization method. Also, the values of $z_{t}$ and $j_{0}$ are given for the best fit model.}
\begin{center}
\begin{tabular}{|c|c|c|c|c|c|c|c|c|}
\hline
$\alpha$&$q_{0}$&$q_{1}$& $\chi^{2}_{m}$&$z_{t}$&$j_{0}$&${\bar{\chi}}^{2}$\\
&($1\sigma$)&($1\sigma$)&&($1\sigma$)&($1\sigma$)&\\
\hline
&&&&&&\\
$1$&$-0.537^{+0.109}_{-0.11}$&$1.276^{+0.276}_{-0.281}$&$32.768$&$0.726^{+0.036}_{-0.041}$&$1.315^{+0.175}_{-0.184}$&$0.819$\\
&&&&&&\\
\hline
&&&&&&\\
$0.5$&$-0.503^{+0.103}_{-0.096}$&$1.008^{+0.214}_{-0.225}$&$32.435$&$0.775^{+0.038}_{-0.039}$&$1.011^{+0.131}_{-0.135}$&$0.811$\\
&&&&&&\\
\hline
&&&&&&\\
$0.3$&$-0.489^{+0.102}_{-0.092}$&$0.911^{+0.191}_{-0.207}$&$32.334$&$0.795^{+0.031}_{-0.046}$&$0.9^{+0.114}_{-0.105}$&$0.808$\\
&&&&&&\\
\hline
&&&&&&\\
$0.007$&$-0.466^{+0.093}_{-0.096}$&$0.774^{+0.169}_{-0.165}$&$32.219$&$0.828^{+0.037}_{-0.053}$&$0.742^{+0.106}_{-0.094}$&$0.805$\\
&&&&&&\\
\hline
&&&&&&\\
$-0.5$&$-0.433^{+0.086}_{-0.088}$&$0.581^{+0.132}_{-0.125}$&$32.139$&$0.884^{+0.061}_{-0.01}$&$0.522^{+0.084}_{-0.081}$&$0.803$\\
&&&&&&\\
\hline
&&&&&&\\
$-1$&$-0.410^{+0.085}_{-0.08}$&$0.437^{+0.093}_{-0.098}$&$32.113$&$0.938^{+0.036}_{-0.05}$&$0.363^{+0.053}_{-0.049}$&$0.802$\\
&&&&&&\\
\hline
\end{tabular}
\label{table1}
\end{center}
\end{table*}
\section{Conclusions}\label{conclusion}
In this paper, we have studied the accelerated expansion phase of the universe by using the kinematic
approach. In this context, we have parameterized the deceleration parameter $q$ in a model independent manner. The functional form of $q$ is chosen in such a way that it reproduces three popular $q$-parametrized models, such as  $q\propto z$, $q\propto {\rm ln}(1+z)$ and $q\propto \frac{z}{1+z}$ for $\alpha=-1$, $\alpha\rightarrow 0$ and $\alpha=1$ respectively. Consequently, the jerk parameter $j$ also incorporates a wide class of viable models of cosmic evolution based on the choice of the parameter $\alpha$. We have also constrained the model parameters by $\chi^{2}$ minimization technique using the latest 41 points of $H(z)$ dataset. Figure \ref{figc} shows the $1\sigma$ and $2\sigma$ confidence level contours in the $q_{0}$-$q_{1}$ parametric plane for different choices of $\alpha$. The best fit values of the model parameters, transition redshift $z_{t}$ and $j_{0}$ within $1\sigma$ error regions for different values of $\alpha$ are displayed in the table \ref{table1}. The ${\bar{\chi}}^2$ implies the same goodness of the model for all the values of $\alpha$ considered here. We have also shown the evolution of the normalized Hubble parameter for our model and have compared that with the latest $H(z)$ dataset. In what follows, we
have summarized our main conclusions in more detail:\\ 
\par For all the values of $\alpha$ considered here, the kinematic model shows a smooth transition from the deceleration ($q>0$) phase to acceleration ($q<0$) phase of the universe in the recent past. It has been found that values of the transition redshift $z_{t}$ (from decelerated to accelerated expansion) and $q_{0}$ depend upon the choice of $\alpha$ . However, the changes in the values of $z_{t}$ and $q_{0}$ do not differ by very large values. It has also been found that the values of $z_{t}$ obtained in this work for a wide range in the values of $\alpha$, are in good agreement with the previous results as reported in  \cite{jerk4,fr2013,capo2014,maga2014,nair2012,mamon2016a,mamon2017a,mamon2017b,fr2017}. We have found from figure \ref{figj} and table \ref{table1} that the $\Lambda$CDM model is not well supported within $1\sigma$ confidence level at the present epoch, except for the cases $\alpha=0.5$ and $0.3$. As discussed earlier, the present model is allowed to pick up any values of $j$ depending on the parameters to be fixed by the observed data as contrary to the work of Zhai et al. \cite{zhaij}, where $j$ is constrained to mimic a flat $\Lambda$CDM model at $z=0$. It has also been found that our model is well consistent with the $H(z)$ data at the low redshifts for different choices of $\alpha$ (figure \ref{figh}). Therefore, we conclude the present $H(z)$ data provides well constrained values of $j$ and our model remains at a very close proximity of the standard $\Lambda$CDM model. However, it is natural to extend the present work with addition of Hubble parameter dataset from the GW standard sirens of neutron star binary system \cite{gwssm}.
\section{Acknowledgments} 
AAM acknowledges the financial support from the Science and Engineering Research Board (SERB), Government of India through National Post-Doctoral Fellowship Scheme (File No: PDF/2017/000308). AAM also wishes to thank the Inter University Center for Astronomy and Astrophysics (IUCAA), Pune for their warm hospitality as a part of the work was done during a visit. The work of KB was supported in part by the JSPS KAKENHI Grant Number JP25800136 and Competitive Research Funds for Fukushima University Faculty (17RI017).  

\end{document}